\documentclass[11pt,letterpaper]{article}

\usepackage[margin=1in]{geometry}
\usepackage{amsmath,amssymb,amsfonts}
\usepackage{graphicx}
\graphicspath{{paper/}}
\usepackage[numbers,sort&compress]{natbib}
\usepackage{booktabs}
\usepackage{xcolor}
\usepackage{microtype}
\usepackage{float}
\usepackage{caption}
\usepackage{subcaption}
\usepackage{setspace}
\usepackage{authblk}
\usepackage{hyperref}

\hypersetup{colorlinks=true, linkcolor=blue!60!black, citecolor=blue!60!black, urlcolor=blue!60!black}
\newcommand{\GP}{\mathcal{GP}}
\newcommand{\KL}{\mathrm{KL}}
\newcommand{\E}{\mathbb{E}}
\newcommand{\R}{\mathbb{R}}
\newcommand{\bx}{\mathbf{x}}
\newcommand{\bZ}{\mathbf{Z}}
\newcommand{\bz}{\mathbf{z}}
\newcommand{\bu}{\mathbf{u}}
\newcommand{\bm}{\mathbf{m}}

\title{\textbf{A single inference process underlies the behavior and neural circuits of working memory}}

\author{Dongyu Gong (dongyu.gong@yale.edu)}
\author{Mario Belledonne (mario.belledonne@yale.edu)}
\author{Ilker Yildirim (ilker.yildirim@yale.edu)}
\affil{Yale University}

\date{}

\begin{document}
\maketitle

\begin{abstract}
Working memory holds past inputs online for future behavior. Despite decades of research, accounts of working memory remain divided across levels of analysis, with long-standing debates on whether its capacity reflects discrete slots or continuous resources, how attention improves recall, and whether its representations rely on persistent neural firing or activity-silent states. Here we show that MemGP, a theory that is specified at the computational, algorithmic, and circuit levels of analysis simultaneously, can explain diverse behavioral and neural phenomena in working memory. At the computational level, MemGP casts working memory as probabilistic inference over a continuous function on feature space, with a variational objective that balances memory strength against a Gaussian-process prior. At the algorithmic level, writing, storing, and reading of memories correspond to approximating this posterior with a finite set of inducing points, the resource that all items share. In simulations, MemGP reproduces load-dependent precision loss, attentional benefits during maintenance, and inter-item attraction and repulsion, in each case revealing multiple complex computational dependencies. At the circuit level, the algorithm analytically (without training) maps onto a recurrent circuit, including its architecture and dynamics. Beyond reproducing neural phenomena such as delay-dependent drift, this analytical map deduces center--surround connectivity, a property long assumed in attractor models, and a structured combination of persistent activity and synaptic weights to hold memories. 
Thus, a single inference objective not only jointly predicts the behavioral and neural signatures of working memory, but also addresses, across several instances and different levels of analysis, existing competing explanations.
\end{abstract}

\section{Introduction}

Human behavior often depends on information that is no longer present in the sensory input, and working memory retains this information for short periods. Despite this centrality to cognition and decades of research, multiple aspects of working memory remain debated. Set-size-dependent limits on recall have been attributed to a fixed number of discrete slots~\cite{luckCapacityVisualWorking1997c,zhangDiscreteFixedresolutionRepresentations2008b} or a continuous resource~\cite{baysPrecisionVisualWorking2009b,maChangingConceptsWorking2014}. Goal-directed attention can improve recall after encoding has ended~\cite{griffinOrientingAttentionLocations2003,van2023turning}, and this benefit has been attributed to the cued memory being strengthened~\cite{liPrioritizingFlexibleWorking2023,baysNeuralModelRetrospective2018a} or protected from decay~\cite{matsukuraAttentionEffectsVisual2007,pertzovRapidForgettingPrevented2013}. Retention across the delay has been attributed to persistent neural firing~\cite{wang2001synaptic,bouchacourt2019flexible} or activity-silent states~\cite{mongilloSynapticTheoryWorking2008,wolffDynamicHiddenStates2017}. Harder still, these debates span the levels at which a cognitive capacity can be analyzed: the computational problem it must solve, the algorithm and representation that realize a solution, and the neural circuits that implement the algorithm~\cite{marrVision1982,dayan2005theoretical}. Existing accounts of working memory mostly address these levels separately. Studies that relate these levels~\cite{weiDistributedResourcesLimited2012,baysNoiseNeuralPopulations2014a,schneegansStochasticSamplingProvides2020,schurginPsychophysicalScalingReveals2020} generally connect only a pair and target mainly the capacity limit. A theory that connects all three levels and explains the broad phenomena of working memory beyond the capacity limit remains a challenge out of the reach of current theories.

Here we propose MemGP, a theory of working memory that addresses this challenge. We start from the computation that working memory performs and derive the algorithm and the circuit from it. At the computational level, MemGP infers a posterior distribution over a memory-strength function on a continuous feature space, using a variational objective that rewards agreement with the evidence and penalizes departure from a Gaussian-process (GP) prior~\cite{williams2006gaussian}. This posterior stores content and uncertainty jointly~\cite{knillBayesianBrainRole2004,liJointRepresentationWorking2021}. At the algorithmic level, a finite set of inducing points~\cite{titsiasVariationalLearningInducing2009}, the resource that all items share, approximates the posterior. Encoding (writing) and maintenance (storing) update this approximation by optimizing the objective, and retrieval (reading) samples from it. At the circuit level, the algorithm maps analytically, without training, onto a recurrent circuit, where gradient descent on the variational objective determines the dynamics of the network.

In simulations of a continuous-reproduction task~\cite{wilkenDetectionTheoryAccount2004,zhangDiscreteFixedresolutionRepresentations2008b,baysPrecisionVisualWorking2009b}, MemGP reproduces load-dependent precision loss, attentional benefits during maintenance, and inter-item attraction and repulsion. The by-construction interpretability of MemGP allowed us to localize each effect to a component of the posterior, in addition to predicting complex dependencies between task parameters and behavioral phenomena. For example, we find that goal-directed attention, implemented as a reweighting of the maintenance objective, acts mainly on the uncertainty of the cued memory: its uncertainty falls more than that of uncued memories, whereas its expected strength declines over the delay with no significant difference from uncued memories. This result provides a novel account of attention in working memory that differs from existing perspectives suggesting strengthening of the cued memory~\cite{liPrioritizingFlexibleWorking2023,baysNeuralModelRetrospective2018a} or protection of its strength from decay~\cite{matsukuraAttentionEffectsVisual2007,pertzovRapidForgettingPrevented2013}. In the recurrent circuit derived from MemGP, the represented feature drifts with delay, which resembles delay-dependent drift reported in prefrontal recordings~\cite{wimmerBumpAttractorDynamics2014}; moreover, center--surround connectivity, a convenient modeling choice made in attractor models~\cite{compteSynapticMechanismsNetwork2000}, follows from this analytical mapping of the algorithm onto the circuit. The circuit holds a memory in a structured combination of persistent activity, which represents expected strength, and a rapidly updated state, interpretable as fast synaptic weights, which represents uncertainty. A single inference process thus connects the behavioral and neural signatures of working memory and bears on each debate: a discrete resource supports continuous precision, attention during maintenance changes mainly the uncertainty of a memory, and persistent activity and fast synaptic weights represent different components of the posterior.

\section{MemGP framework}

MemGP specifies how perceptual evidence is written into a memory, how that memory is maintained during the delay, and how it is read out for report. Encoding forms a posterior over memory-strength functions from sensory evidence, maintenance rehearses that posterior without new sensory input, and retrieval samples the posterior to generate a color report.

\begin{figure}[p]
    \centering
    \includegraphics[width=\textwidth,height=0.66\textheight,keepaspectratio]{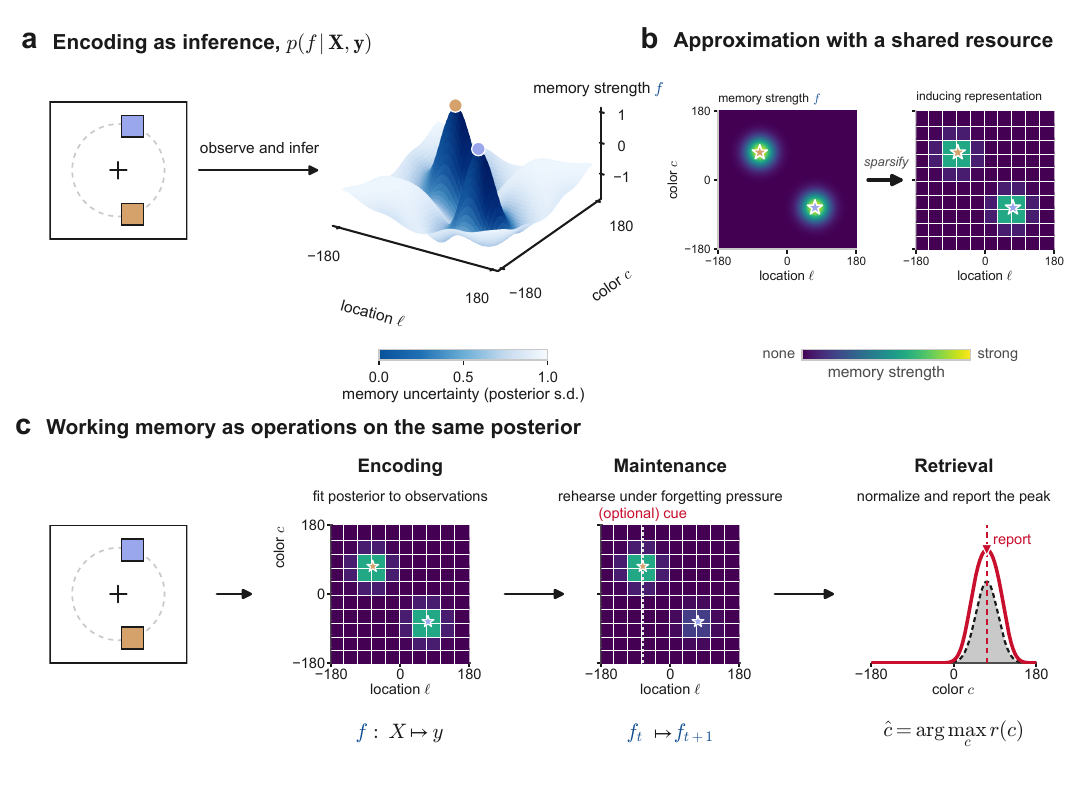}
    \caption{\textbf{The MemGP framework.}
    \textbf{(a)} Encoding as inference. Two colored items are observed and inferred as a GP posterior over the location$\,\times\,$color torus. Surface height is the posterior mean, with a localized peak at each item, and color indicates memory uncertainty (posterior standard deviation).
    \textbf{(b)} A finite inducing representation supports a continuous function. The schematic shows memory strength across the feature space and at $M=100$ inducing points. The regular grid illustrates initialization; inducing positions are updated in the behavioral simulations. Color indicates memory strength.
    \textbf{(c)} Encoding infers the posterior from perceptual evidence; maintenance uses the preceding posterior mean as rehearsal targets, with selective attention increasing the likelihood weights near the cued location and KL regularization penalizing departure from the prior; retrieval draws a joint posterior sample, divisively normalizes the probed response by the pool of all items, and reports $\arg\max_c r(c)$. Grey fill: normalization pool; black curve: probed response $f_t^{+}$; red curve: normalized readout $r(c)$. The stage illustrations are schematic; full expressions are given in Eqs.~\ref{eq:elbo_enc}--\ref{eq:elbo_maint} and Supplementary S1.5.}
    \label{fig:schematic}
\end{figure}

\subsection{Memory strength and uncertainty}

Memory strength is a latent function $f$ on the feature space (Fig.~\ref{fig:schematic}a).
Here, we consider the continuous-reproduction task~(Fig.~\ref{fig:setsize}a), in which an array of $N$ colored items appears briefly. After a blank delay, a location probe identifies an item whose color must be reported on a continuous wheel.
So, each input to the latent memory strength function $f$ is a location--color pair $\bx = (\ell, c)$; the value $f(\bx)$ is a memory-strength field: how strongly the system holds that an item of color $c$ appeared at location $\ell$. We place a GP prior over $f$ with a product of periodic kernels matching each dimension's circular geometry:
\begin{equation}
    f \sim \GP\bigl(m(\bx),\; k(\bx, \bx')\bigr)
    \label{eq:prior}
\end{equation}
where $m(\bx)$ is a constant-valued mean function, $k(\bx, \bx')$ is a periodic kernel with period $360^\circ$ and separate lengthscale parameters for location, $\lambda_l$, and color, $\lambda_c$. Under this kernel, nearby location--color combinations have more strongly correlated memory strengths than distant combinations (Supplementary Fig.~\ref{fig:prior}). Evidence about an observed color at one location can therefore inform predicted strength at nearby colors and locations. The memory representation is the GP posterior over possible strength functions (Fig.~\ref{fig:schematic}a), which stores both the content and its uncertainty. At each location--color combination, the posterior mean specifies expected strength and the posterior variance specifies uncertainty about that strength.

\subsection{Inducing points as a shared computational resource}

MemGP approximates the posterior over memory-strength functions using $M$ inducing points, $\bZ=\{(\ell_m^{(z)},c_m^{(z)})\}_{m=1}^{M}$, and their function values $\bu=f(\bZ)$. The inducing points begin on a regular grid and can move during both encoding and maintenance (Fig.~\ref{fig:schematic}b). A mean vector $\bm\in\R^M$ specifies expected memory strength at the inducing points. A covariance matrix $\mathbf{S}\in\R^{M\times M}$ specifies uncertainty and dependencies among these strengths. Together they define the Gaussian variational distribution~\cite{titsiasVariationalLearningInducing2009}
\begin{equation}
q(\bu)=\mathcal{N}(\bm,\mathbf{S}).
\label{eq:qu}
\end{equation}
For a queried location--color combination $\bx^*$, the GP conditional specifies how its strength depends on the inducing values. Integrating over the uncertainty in those values gives the approximate posterior:
\begin{equation}
q(f(\bx^*))=\int p(f(\bx^*)\mid\bu)q(\bu)d\bu.
\label{eq:q}
\end{equation}

The inducing points provide a finite resource shared by all items. Adding another item does not add inducing points, so the model must represent more location--color combinations using the same number of inducing values. Kernel interpolation predicts memory strength at feature combinations between inducing points, so remembered colors and locations need not coincide with the inducing locations. An item generally depends on several inducing points, and each inducing point can contribute to several items. The computational resource of MemGP is thus discrete while the precision of each item varies continuously.

\subsection{Encoding, maintenance and retrieval}

During encoding, a procedure that approximates sensory processing supplies noisy location--color observations $\mathbf{X}=\{\bx_i\}_{i=1}^{n}$ and evidence strengths $\mathbf{y}=\{y_i\}_{i=1}^{n}$, with larger strengths for observations closer to an item's location and color (Fig.~\ref{fig:schematic}c; see Methods). A Gaussian likelihood $p(y_i\mid f_i)$ relates this evidence to the latent memory strength $f_i=f(\bx_i)$. To infer the latent memory strengths from these observations, variational inference adjusts the approximate posterior by optimizing an objective called the evidence lower bound (ELBO). During encoding, MemGP minimizes its negative:
\begin{equation}
    \mathcal{L}_{\mathrm{enc}}=-\sum_{i=1}^{n}\E_{q}\!\left[\log p(y_i\mid f_i)\right]+\KL\bigl(q(\bu)\,\|\,p(\bu)\bigr).
    \label{eq:elbo_enc}
\end{equation}
The expected log-likelihood term rewards a posterior that predicts the observed evidence strengths. The Kullback--Leibler (KL) divergence measures how far the posterior over inducing values departs from their GP prior. Minimizing the negative ELBO therefore balances agreement with the observations against a penalty for departing from the prior. Because the posterior uses a fixed number of inducing points, all items must be represented within the same finite approximation.

During the delay, external observations are replaced by the model's own previous predictions. The model maintains the memory representation by minimizing:

\begin{equation}
    \mathcal{L}_{\mathrm{maint}} = -\frac{1}{M}\sum_{m=1}^{M} a(\ell^{(z)}_m) \, \E_q\!\left[\log p(\tilde{y}_m \mid f(\bz_m))\right] + \beta \, \KL\bigl(q(\bu) \,\|\, p(\bu)\bigr)
    \label{eq:elbo_maint}
\end{equation}

\noindent where $\bz_m = (\ell^{(z)}_m, c^{(z)}_m)$ is the position of inducing point $m$ at epoch $t$, and $\tilde{y}_m = \E_{q}[f(\bz_m)]\big|_{t-1}$ is the posterior mean strength at that point from the previous epoch $t-1$, which serves as the rehearsal target. The rehearsal term favors predictions that remain close to the preceding epoch's memory strengths. The KL term mathematically grounds forgetting over time, penalizing departure from the GP prior with relative weight $\beta$ (weighting the KL term is common in variational inference, e.g., in~\cite{higgins2017beta}). The attention weight $a(\ell^{(z)}_m)$ modulates the memory precision near the region $\ell^{(z)}_m$ (Eq.~\ref{eq:attention}). In the continuous-reproduction task, attentional cues are the spatial location $\check{\ell}_j$ of a target item $j$, and the weights are a Gaussian spatial gain (Fig.~\ref{fig:retrocue}b) depending only on the distance between $\check{\ell}_j$ and the inducing point's location $\ell^{(z)}_m$. When no cue is presented, all weights equal~$1$.

At retrieval, MemGP draws one joint posterior sample of memory strength across colors at all occupied item locations. The sampled strength functions are first made nonnegative by subtracting their global minimum. A spatial divisive-normalization rule~\cite{carandiniSummationDivisionNeurons1994,carandiniNormalizationCanonicalNeural2012} then divides the nonnegative strength function at the probed location by a baseline term plus a weighted pool of strength functions across remembered locations (Supplementary Eq.~\ref{eq:pool}). Items spatially closer to the probe contribute more to the pool (Supplementary Eq.~\ref{eq:pool_weight}). The model reports the color with the largest normalized response, $\hat c=\arg\max_c r(c)$ (Supplementary S1.5, Eq.~\ref{eq:readout}). Thus, encoding and maintenance determine the posterior over strength functions, while sampling and normalization convert that posterior into a color report.

\section{Results}

\subsection{Shared capacity produces load-dependent errors}

\begin{figure}[p]
    \centering
    \includegraphics[width=\textwidth,height=0.72\textheight,keepaspectratio]{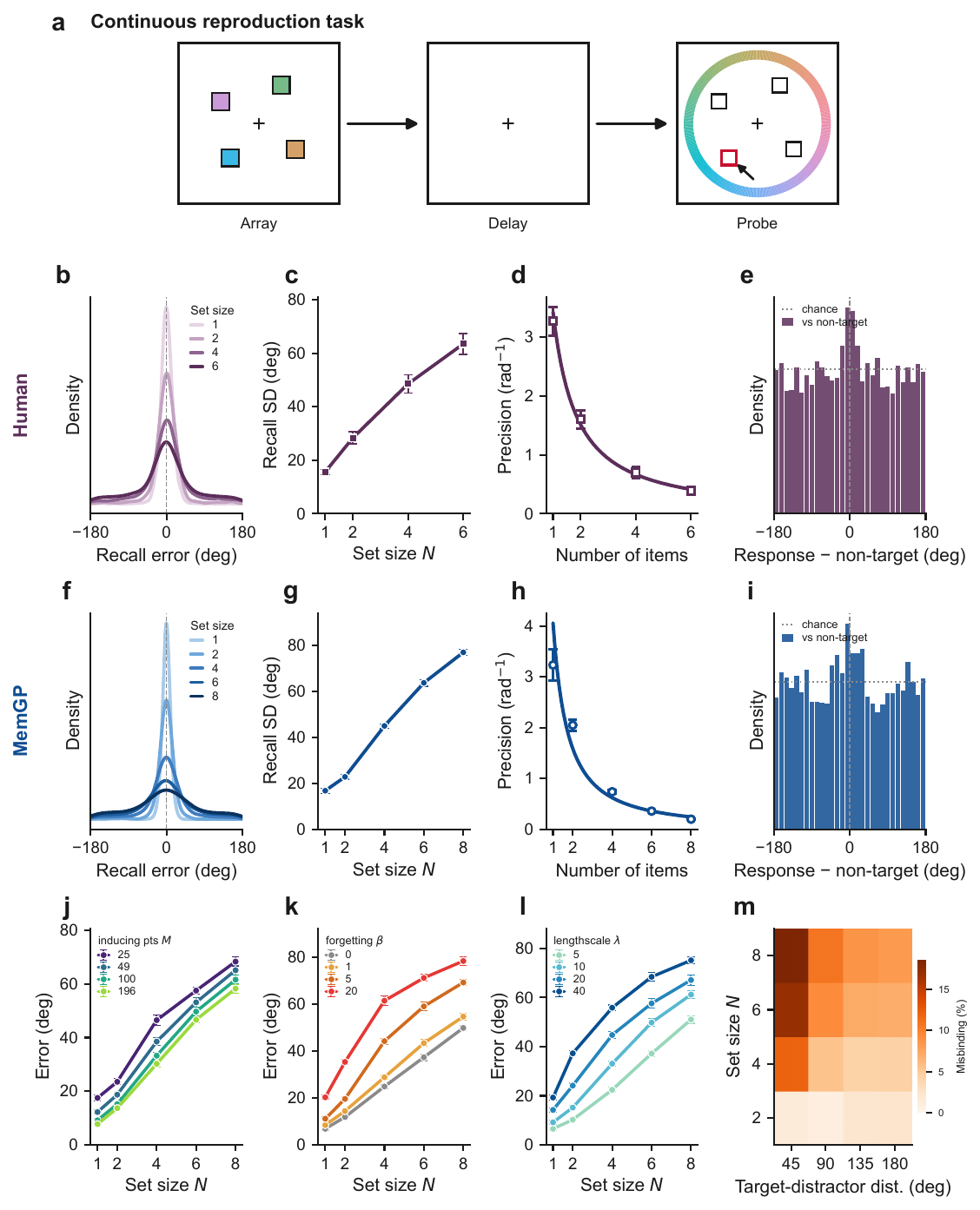}
    \caption{\textbf{Capacity limits emerge from a shared finite representation.}
    \textbf{(a)} Continuous-reproduction task: stimulus array, blank delay, and color-wheel probe at one location.
    \textbf{(b--e)} Human benchmark behavior. \textbf{(b,c)} Color error distributions and circular SD vs.\ set size, as in \citet{maChangingConceptsWorking2014}(underlying trials from \citet{baysPrecisionVisualWorking2009b}). \textbf{(d)} Precision (inverse circular SD) vs.\ set size with power-law fit, from \citet{baysPrecisionVisualWorking2009b}. \textbf{(e)} Distribution of reported-minus-non-target color differences at $N\!=\!4$, from \citet{baysPrecisionVisualWorking2009b}.
    \textbf{(f--i)} MemGP error distributions, circular SD, precision and non-target-relative reports, in the same order as the human panels.
    \textbf{(j--l)} Error vs.\ set size across inducing-grid sizes $M \in \{25, 49, 100, 196\}$ (j), KL weights $\beta \in \{0, 1, 5, 20\}$ (k), and initial kernel lengthscales $\lambda \in \{5, 10, 20, 40\}$ with location and color kernels set equal (l). Each changes recall error magnitude and preserves the monotonic increase in recall error with set size.
    \textbf{(m)} Misbinding rate (probability that the probe is reported as a non-target item) as a joint function of set size and target--distractor spatial distance: swap errors fall with distance and rise with load.
    In \textbf{(g, h, j--l)}, points are means $\pm$ s.e.m.\ across 20 simulated subjects, with one item probed per trial (100 trials per set size in g,h; 50 in j--l). In \textbf{(m)}, each cell is the misbinding rate averaged over all 20 subjects.}
    \label{fig:setsize}
\end{figure}

We first tested whether increasing the number of remembered items reduces recall precision when all items share the same inducing points. Across set sizes $N\in\{1,2,4,6,8\}$, MemGP reproduced canonical signatures of human continuous-report performance (Fig.~\ref{fig:setsize}b--i). As load increased, the error distributions broadened (Fig.~\ref{fig:setsize}b,f;~\cite{wilkenDetectionTheoryAccount2004,zhangDiscreteFixedresolutionRepresentations2008b,baysPrecisionVisualWorking2009b}) and their circular standard deviation (SD) increased (Fig.~\ref{fig:setsize}c,g;~\cite{wilkenDetectionTheoryAccount2004,zhangDiscreteFixedresolutionRepresentations2008b,baysPrecisionVisualWorking2009b}). Precision, defined as inverse circular SD, declined approximately as a power law, $1/\mathrm{SD}\propto N^{-\alpha}$ (Fig.~\ref{fig:setsize}d,h;~\cite{baysPrecisionVisualWorking2009b}). Reports also clustered above chance around the colors of other items in the display, producing swap errors (`misbinding';~\cite{baysPrecisionVisualWorking2009b}) in which a non-target color is reported in response to the target location (Fig.~\ref{fig:setsize}e,i).
    
MemGP reproduced these patterns without assumptions about the form of the error distribution. Descriptive models \cite{zhangDiscreteFixedresolutionRepresentations2008b, baysPrecisionVisualWorking2009b, maChangingConceptsWorking2014,vandenbergVariabilityEncodingPrecision2012} specify the error distribution directly, for example as a mixture of von Mises components centered on the target and non-target colors and a uniform component \cite{baysPrecisionVisualWorking2009b}, and they leave the mechanism that produces the errors unspecified. In MemGP, the error distribution is an outcome of inference, sampling and normalization. The results are robust to the choice of hyperparameters, which change the magnitude of recall error but not its monotonic increase with set size: a larger inducing grid $M$ lowered error at every set size (Fig.~\ref{fig:setsize}j), and a larger KL weight $\beta$ (Fig.~\ref{fig:setsize}k) or a wider kernel lengthscale $\lambda$ (Fig.~\ref{fig:setsize}l) raised it at every set size. Set-size effects in MemGP therefore follow from items competing for a finite set of inducing points under the inference process that governs encoding and maintenance.

Because MemGP specifies a mechanism, it can be used to examine interactions that descriptive models do not address, such as the joint dependence of swap errors on set size and target--distractor similarity. In simulations that varied both factors, the probability of misbinding the probe to a non-target decreased with target--distractor distance in the cue dimension, here spatial location, and increased with set size (Fig.~\ref{fig:setsize}m), consistent with human reports~\cite{baysPrecisionVisualWorking2009b,emrichCompetitionIncreasesBinding2012,rerkoSpatialTranspositionGradients2014}. A graded effect of target--distractor similarity is a signature of feature-binding errors~\cite{baysEvaluatingExcludingSwap2016,schneegansNeuralArchitectureFeature2017c,mcmasterSwapErrorsVisual2022}, which population-coding models produce through dedicated binding mechanisms such as conjunction-selective units~\cite{schneegansNeuralArchitectureFeature2017c}. MemGP requires no separate binding rule. The kernel correlates memory strengths at nearby locations, so evidence for a neighboring item's color enters the strength profile at the probed location and can determine the reported color.

Encoding precision in human working memory varies across items and trials~\cite{vandenbergVariabilityEncodingPrecision2012,fougnieVariabilityQualityVisual2012}. The variable-precision (VP) model captures this variability by assuming that precision follows a Gamma distribution across items and trials, and it fits human data better than models with fixed precision~\cite{vandenbergVariabilityEncodingPrecision2012}. However, the VP model does not specify why precision varies. MemGP has no parameter for precision variability, yet precision variability arises from posterior uncertainty. Following \citet{vandenbergVariabilityEncodingPrecision2012}, we fitted uniform--von~Mises mixtures~\cite{zhangDiscreteFixedresolutionRepresentations2008b} to MemGP's errors at each set size. We then compared the recovered mixture weight $w$ and circular SD with the predictions of the item-limit, slots-plus-averaging, equal-precision and VP models, each fitted to the same errors (Supplementary Fig.~\ref{fig:model_comparison}). The VP model had the lowest Akaike information criterion (AIC) and Bayesian information criterion (BIC) in all 20 simulated subjects (Supplementary Table~\ref{tab:ic}) and the lowest error on both mixture statistics ($\mathrm{RMSE}_w=0.092$; $\mathrm{RMSE}_{\mathrm{CSD}}=0.141$). The trial-by-trial and item-by-item variability in MemGP's precision therefore has a form similar to what the VP model assumes. Posterior uncertainty under capacity-limited inference is thus a candidate mechanism for the precision variability. 

\subsection{Selective maintenance reduces posterior variance and improves recall}

\begin{figure}[p]
    \centering
    \includegraphics[width=\textwidth,height=0.68\textheight,keepaspectratio]{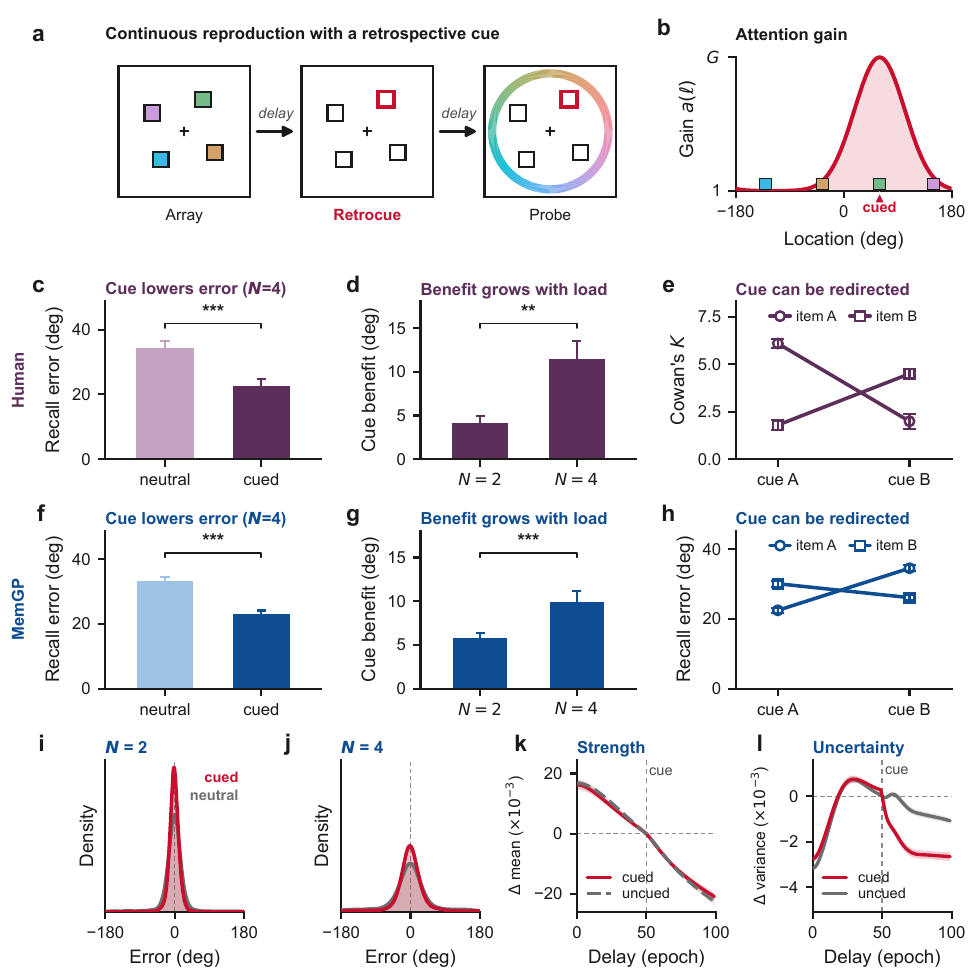}
    \caption{\textbf{Attention is a reweighting of the maintenance objective.}
    \textbf{(a)} Continuous reproduction with a retrospective cue: array, delay, a retrocue indicating one item, delay, and probe.
    \textbf{(b)} Spatial attention gain $a(\ell)$: a Gaussian gain profile centered on the cued location, equal to $G$ at peak and to~$1$ away from the cue (Eq.~\ref{eq:attention}). Squares on the location axis mark the locations of the four items.
    \textbf{(c--d)} Human continuous-report data from \citet{souzaFocusedAttentionImproves2014}: lower absolute error for focused than neutral items at $N\!=\!4$ (c); cue benefit (neutral minus cued absolute recall error) increasing with set size (d).
    \textbf{(e)} Human cue-redirect pattern from \citet{vanMoorselaarForgottenButNot2015}, Exp.~2: Cowan's $K$ for items A and B after an initial cue to A and after a mid-delay redirect to B (change-detection capacity; digitized from their Fig.~2b).
    \textbf{(f--h)} MemGP. Cued error is below neutral, and the benefit is larger at $N\!=\!4$ than at $N\!=\!2$ (f--g). After a mid-delay redirect, error rises for item A and falls for item B (h).
    \textbf{(i--j)} Recall-error distribution, neutral vs.\ cued, at $N\!=\!2$ (i) and $N\!=\!4$ (j): cueing concentrates the recall-error distribution more strongly around zero, with a larger effect at higher load.
    \textbf{(k)} Change in cued vs.\ uncued memory strength over the delay (cue at the vertical dashed line): posterior mean declines in both conditions, and the difference between their changes is not significant.
    \textbf{(l)} Change in posterior variance over the same delay, cued item versus the other items: the decrease is larger for the cued item.
    Bars and points in \textbf{(c--h)} are means $\pm$ s.e.m. In \textbf{(f--h)} there are 20 simulated subjects, 100 trials per condition. Traces in \textbf{(k--l)} are means $\pm$ s.e.m.\ across 100 trials.}
    \label{fig:retrocue}
\end{figure}

Working memory is sensitive to retrospective attentional cues (retrocues) presented during the retention interval(Fig.~\ref{fig:retrocue}a), which provide benefits for the recall performance of cued items~\cite{griffinOrientingAttentionLocations2003,souzaRetrocuesRetroactiveAttentional2016,van2023turning}. 
In MemGP, a retrospective cue increases the likelihood weights of inducing points near the cued item in the maintenance objective (Eq.~\ref{eq:attention}; Fig.~\ref{fig:retrocue}b). The increased weights give memory strengths near the cued location greater influence on rehearsal. Neutral trials retain equal weights at all inducing points. A broad range of human performance patterns fall from this reweighted maintenance in MemGP: lower recall error for cued items, a larger benefit at higher load, and a transfer of the benefit when a second cue selected a different item (Fig.~\ref{fig:retrocue}c--e;~\cite{souzaFocusedAttentionImproves2014,vanMoorselaarForgottenButNot2015,huLoadDependentRetrospectiveAttentional2025}). We describe them in turn. 

\textbf{Load-dependent benefit.} In MemGP, cued items were recalled more accurately than neutral items at both set sizes (Fig.~\ref{fig:retrocue}f). At $N=2$, mean absolute error was $9.2^\circ$ for cued items and $15.1^\circ$ for neutral items, a benefit of $5.9^\circ\pm0.4^\circ$ (paired $t(19)=13.39$, $p<0.001$). At $N=4$, the corresponding errors were $23.4^\circ$ and $33.4^\circ$, a benefit of $10.0^\circ\pm1.1^\circ$ (paired $t(19)=8.82$, $p<0.001$). The benefit was $4.1^\circ$ larger at $N=4$ than at $N=2$ ($t(38)=3.40$, $p<0.001$, one-tailed; Fig.~\ref{fig:retrocue}g). Selective maintenance in MemGP therefore had a greater effect when more items shared the representation, consistent with load-dependent cue benefits in humans~\cite{souzaFocusedAttentionImproves2014,huLoadDependentRetrospectiveAttentional2025}.

\textbf{Flexible reallocation.} When a second cue redirected attention from item A to item B during the delay, MemGP's error rose by $12.1^\circ$ for A and fell by $4.0^\circ$ for B (Fig.~\ref{fig:retrocue}h). The difference between the error changes for the two items, $\Delta\mathrm{A}-\Delta\mathrm{B}$, was $16.1^\circ$ (95\% CI $[13.3,18.8]$; paired $t(19)=12.24$, $p<0.001$). The loss of precision for the previously cued item and improvement for the newly cued item agree in direction with the double-cue results reported by \citet{vanMoorselaarForgottenButNot2015} (Fig.~\ref{fig:retrocue}e). Their experiment measured change-detection capacity, whereas the MemGP simulations measured continuous-report error, so the comparison concerns the direction of the effect across different behavioral measures (Cowan's $K$ vs. recall error).

\textbf{Maintenance dynamics.} Cueing concentrated the recall-error distribution more strongly around zero, with a larger change at $N=4$ (Fig.~\ref{fig:retrocue}i--j). To relate this benefit to the maintained representation, we tracked the posterior mean and variance of cued and uncued items from cue onset to the end of the delay. If attention strengthened the cued memory~\cite{liPrioritizingFlexibleWorking2023}, its posterior mean would increase after the cue; if attention protected the cued memory from decay~\cite{matsukuraAttentionEffectsVisual2007,pertzovRapidForgettingPrevented2013,baysNeuralModelRetrospective2018a}, its posterior mean would decline less than that of uncued items. The posterior mean, which measures expected memory strength, declined for both cued and uncued items; the difference in their changes from cue onset to delay end was not significant (paired $t(99)=1.41$, $p=0.16$; Fig.~\ref{fig:retrocue}k). In contrast, posterior variance, which measures uncertainty about memory strength, decreased more for the cued item than for the mean of the other items (paired $t(99)=-7.53$, $p<0.001$; Fig.~\ref{fig:retrocue}l). In MemGP, the cue benefit therefore accompanied a selective reduction in uncertainty, without an increase in expected strength or a significant reduction in its decline. This dissociation explains why the model needs both strength and uncertainty: lower recall error can accompany reduced posterior variance even while mean strength declines (Supplementary S3.2).

\subsection{Shared representations and normalization produce attraction and repulsion}

\begin{figure}[p]
    \centering
    \includegraphics[width=\textwidth,height=0.68\textheight,keepaspectratio]{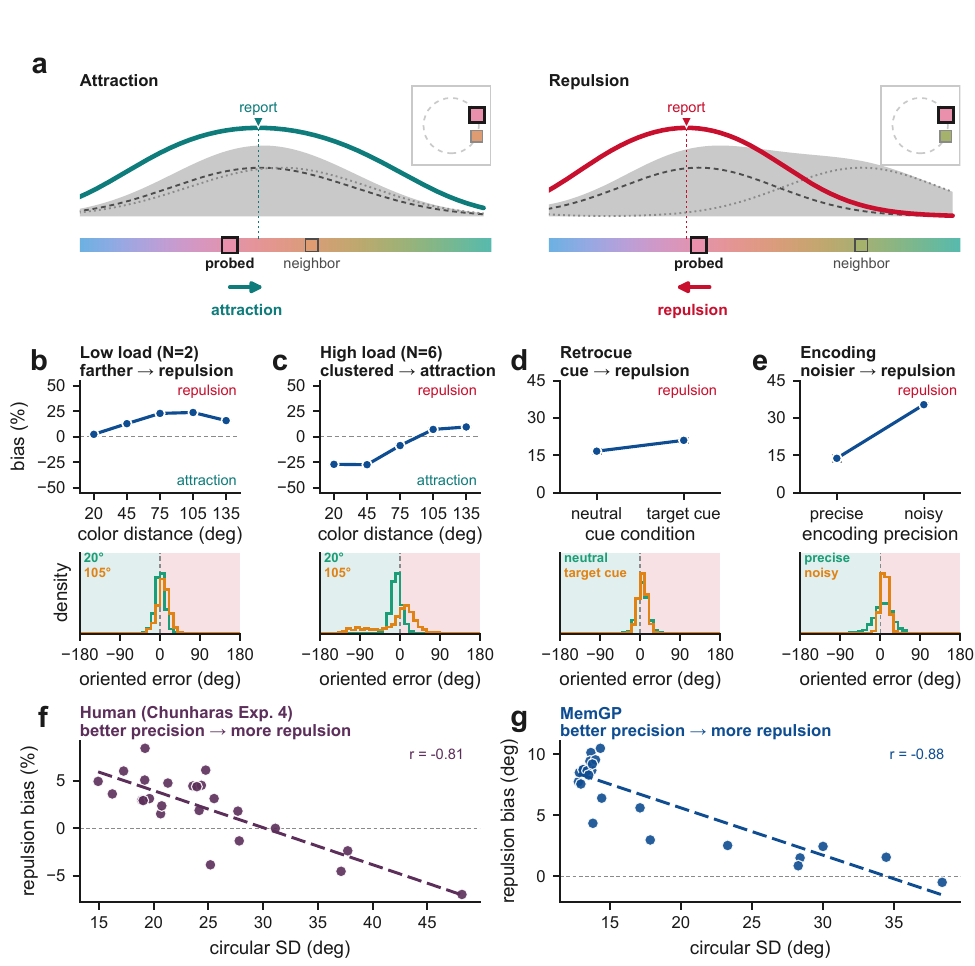}
    \caption{\textbf{Inter-item attraction and repulsion depend on memory demands.}
    \textbf{(a)} Two scenarios for the color profile at the probed location (schematic). Left: a neighbor with a similar color shifts the target profile toward the neighbor; the pool has nearly the same shape, and the readout follows that shift (teal). Right: a neighbor with a more distant color contributes an asymmetric normalization pool that shifts the readout away from the neighbor (red). Dashed curve: target profile; dotted curve: neighbor profile; grey fill: pool; bold: normalized readout.
    \textbf{(b,c)} Bias vs.\ color distance at low load ($N=2$, b) and high load ($N=6$, c); histograms in both panels highlight $20^\circ$ and $105^\circ$.
    \textbf{(d)} A retrocue on the target modestly increases repulsion.
    \textbf{(e)} At color distance $135^\circ$, noisier encoding ($\sigma_c=60^\circ$) increases repulsion relative to precise encoding ($\sigma_c=10^\circ$), matching the Exp.~4 signature in \citet{chunharasAdaptivePerspectiveVisual2022a}. In \textbf{(b--e)}, points are means $\pm$ s.e.m.\ across 20 simulated subjects, 100 trials per condition (s.e.m.\ between $0.8\%$ and $1.4\%$, nearly invisible). Histograms show signed-error distributions (left/right shading: attraction/repulsion).
    \textbf{(f, g)} Lower circular SD predicts stronger repulsion in humans and MemGP. \textbf{(f)} is reproduced from \citet{chunharasAdaptivePerspectiveVisual2022a} Exp.~4 data (24 subjects). In \textbf{(g)}, each MemGP simulated subject ($n=24$) is tested at target--distractor color distances $\{20,40,60,80,100,120,140\}^\circ$; circular SD and signed angular bias are averages across those distances.}
    \label{fig:bias}
\end{figure}

When several items are held in memory, the recalled feature of one item is often biased by its neighbors: the report may shift toward a neighboring feature (attraction)~\cite{bradyHierarchicalEncodingVisual2011} or away from it (repulsion)~\cite{baeInteractionsVisualWorking2017,scottiVisualWorkingMemory2021}. These biases have been interpreted as adaptive~\cite{chunharasAdaptivePerspectiveVisual2022a}. Repulsion can keep a few confusable items distinct, and attraction can stabilize noisy estimates of individual items with information from the ensemble~\cite{bradyHierarchicalEncodingVisual2011,utochkinIndividualRepresentationsVisual2020,livelyMemoryFidelityReveals2021,miaoBridgingBayesianRepresentational2026}. Previous models produce attraction~\cite{swanBindingPoolModel2014a,oberauerInterferenceModelVisual2017a} and, in some cases, repulsion~\cite{almeida2015neural,bouchacourt2019flexible}, but they mainly address how the bias depends on the feature distance between items. In human data, the direction and size of the bias also depend on multiple confounding factors~\cite{chunharasAdaptivePerspectiveVisual2022a}. MemGP accounts for these dependencies through its retrieval rule (Fig.~\ref{fig:bias}a). The sampled strength profile over color at the probed location is divided by a baseline term plus a weighted pool of the sampled profiles of all items in the display. The direction of the bias depends on where the pool lies relative to the target profile. When a neighbor's color is close to the target's, kernel generalization shifts the target profile toward the neighbor, and the pool, formed from the same overlapping profiles, has nearly the same shape over color as the target profile. Division by the pool therefore leaves the peak position largely unchanged, the normalized response follows the shifted profile, and the report is attracted (Fig.~\ref{fig:bias}a, left). When the neighbor's color is farther away, the pool is larger on the side of the target profile that faces the neighbor, so division suppresses that side and the report is repelled (Fig.~\ref{fig:bias}a, right).

We quantified the direction of bias using the model-free measure of \citet{chunharasAdaptivePerspectiveVisual2022a}, which counts reports rather than fitting a distribution to the errors:
\begin{equation}
    \mathrm{bias}=100\times\frac{n_{\mathrm{away}}}{n_{\mathrm{total}}}-50,
    \label{eq:bias}
\end{equation}
where $n_{\mathrm{away}}$ counts reports whose error points away from the non-target color, and $n_{\mathrm{total}}$ counts all reports. Subtracting 50 centers the measure on equal proportions of errors toward and away from the non-target. Positive values indicate repulsion and negative values indicate attraction; values are expressed in percentage points relative to 50\%.

At low load ($N=2$), bias was near zero at the smallest color separation ($2.4\pm1.0\%$ at $20^\circ$) and increased to repulsion, peaking around $75$--$105^\circ$ ($23.8\pm1.1\%$ at $105^\circ$; means $\pm$ s.e.m.\ across 20 simulated subjects; Fig.~\ref{fig:bias}b). At high load ($N=6$), with distractor colors placed on the same side of the target, clustered separations produced attraction ($-27.1\pm0.9\%$ at $20^\circ$). Repulsion returned at wider separations ($7.2\pm1.3\%$ at $105^\circ$; Fig.~\ref{fig:bias}c). The same posterior-sampling and normalization rule therefore generated attraction or repulsion depending on memory load and the arrangement of colors, consistent with the human tendency toward attraction under high load~\cite{chunharasAdaptivePerspectiveVisual2022a}.

We also tested whether retrospective attention and the noise in encoding observations change the direction or magnitude of recall bias. Cueing the target during the delay increased repulsion from $16.7\pm1.1\%$ to $21.1\pm1.1\%$ (Fig.~\ref{fig:bias}d). Noisier encoding (color sampling noise $\sigma_c=60^\circ$) also produced greater repulsion than more precise encoding ($\sigma_c=10^\circ$; $35.4\pm0.7\%$ versus $13.9\pm1.4\%$; Fig.~\ref{fig:bias}e), reproducing the direction of the encoding manipulation in \citet{chunharasAdaptivePerspectiveVisual2022a} and \citet{livelyMemoryFidelityReveals2021}. 

In the human data, lower circular SD was associated with greater repulsion (Pearson $r=-0.81$, 95\% CI $[-0.91,-0.60]$; Fig.~\ref{fig:bias}f). We found the same strong negative correlation in the MemGP simulations (Pearson $r=-0.88$, 95\% CI $[-0.95,-0.75]$; Fig.~\ref{fig:bias}g). For both human participants and MemGP simulations, repulsion biases are thus strongest when recall is more precise.

In summary, attraction and repulsion in MemGP arise from one retrieval rule applied to a shared memory representation. Feature distance, load, attention, encoding noise and recall precision each change the direction or size of the bias, as they do in human reports.

\subsection{Variational inference implemented by recurrent circuit dynamics}

\begin{figure}[!hp]
    \centering
    \includegraphics[width=\textwidth,height=0.67\textheight,keepaspectratio]{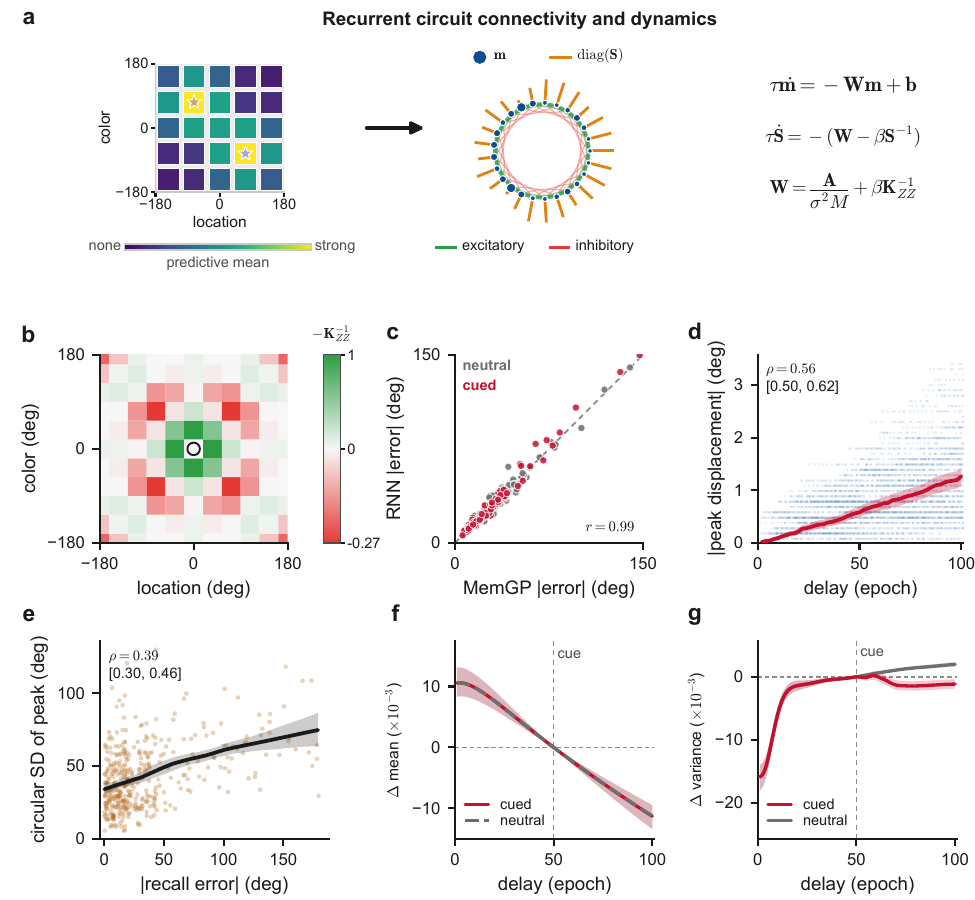}
    \caption{\textbf{Variational updates yield recurrent dynamics with center--surround interactions.}
    \textbf{(a)} A schematic inducing grid ($M=25$; square color is the predictive mean) storing two items is transformed as 25 recurrent units on a ring. Circle size indicates expected strength $\bm$; spoke length indicates marginal variance $\mathrm{diag}(\mathbf{S})$. Green and red arcs indicate excitation and inhibition.
    \textbf{(b)} Outgoing weights $-\mathbf{K}_{ZZ}^{-1}$ from one inducing unit (the open circle; self-coupling removed) on the $10\times10$ grid ($M=100$), computed as $-(\mathbf{K}_{ZZ}+\varepsilon\mathbf{I})^{-1}$ with jitter $\varepsilon=10^{-3}$ (Supplementary Fig.~\ref{fig:kernel_precision}). Weights are scaled by their largest absolute entry; green and red indicate excitation and inhibition.
    \textbf{(c)} Expected recall error (mean absolute error over 100 posterior-sample reports) of MemGP and the recurrent circuit on matched trials. Each point is one trial in the neutral (grey) or retrocue (red) condition. Dashed line: identity. Pearson $r=0.99$ across 200 trials.
    \textbf{(d)} Absolute displacement of the posterior-mean peak from its position at delay onset in the recurrent circuit. Displacement grows with delay under KL regularization ($100$ trials; Spearman $\rho=0.56$).
    \textbf{(e)} Circular SD (over 200 posterior-sample reports) of the posterior-mean peak against the absolute error of an independent report on the recurrent circuit ($100$ trials at $N=4$; Spearman $\rho=0.39$).
    \textbf{(f, g)} Change in posterior mean (f) and variance (g) at the probed item on the recurrent circuit, from their values at cue onset. Each trial was encoded once and then maintained twice, with and without a retrocue ($N=4$, $100$ trials).
    Lines in \textbf{(d, f, g)} are across-trial means; in \textbf{(e)}, points are items and the line is a local-linear fit. Bands are bootstrap 95\% intervals.}
    \label{fig:circuit}
\end{figure}

MemGP accounts for the behavioral effects of memory load, retrospective attention and inter-item interactions. To relate the computations that produce these effects to a circuit, we derived a recurrent circuit analytically from the variational objectives, without training any connection weights. With the inducing locations, prior mean, kernel parameters and likelihood noise held fixed, a memory is specified by the two parameters of $q(\bu)$ (Eq.~\ref{eq:qu}): the mean $\bm$ gives the expected memory strength at each inducing point, and the covariance $\mathbf{S}$ gives the uncertainty about these strengths and the dependencies among them. We map $\bm$ to population activity, with each unit corresponding to a population tuned to a region of location--color space, and $\mathbf{S}$ to a separate, rapidly updated state. The encoding and maintenance objectives (Eqs.~\ref{eq:elbo_enc} and~\ref{eq:elbo_maint}) then specify how these two states change as a memory is formed and held (Fig.~\ref{fig:circuit}a). The Gaussian variational objective is quadratic in the mean (Supplementary Eqs.~\ref{eq:enc_quad} and~\ref{eq:maint_quad}), and taking the gradient of this objective with respect to the mean gives a linear update consisting of recurrent interactions and an input drive (Supplementary Eqs.~\ref{eq:enc_grad} and~\ref{eq:maint_grad}).

For maintenance, the resulting dynamics of the expected memory strength are $\tau\dot{\bm}=-\mathbf{W}\bm+\mathbf{b}$ (Supplementary Eq.~\ref{eq:circuit}), where $\mathbf{W}=\mathbf{A}/(\sigma^2M)+\beta\mathbf{K}_{ZZ}^{-1}$ is the connectivity matrix. The diagonal matrix $\mathbf{A}$ contains the attention gains, $\sigma^2$ is the likelihood noise variance, $M$ is the number of inducing points, and $\beta$ is the maintenance KL weight. The matrix $\mathbf{K}_{ZZ}$ is the prior covariance of the inducing values, so its inverse is the prior precision. The first term of $\mathbf{W}$ comes from attention-weighted rehearsal and sets how strongly each unit is held to its rehearsal target; the second term comes from the KL penalty, which draws the memory toward the prior and produces forgetting. The input drive $\mathbf{b}$ depends on the current rehearsal target and prior mean; $\tau$ sets the timescale of gradient dynamics (Supplementary S3).

The prior-precision term in $\mathbf{W}$, which supplies interactions between units, remains fixed when inducing locations and kernel parameters are fixed. Because the activity equation contains $-\mathbf{W}\bm$, the prior-dependent interactions have the negative signed term $-\mathbf{K}_{ZZ}^{-1}$. 
The off-diagonal weights under this term exhibit local excitation and surrounding inhibition under MemGP's prior (Fig.~\ref{fig:circuit}b), deriving the center--surround (Mexican-hat) pattern assumed in working-memory attractor models~\cite{compteSynapticMechanismsNetwork2000,wimmerBumpAttractorDynamics2014}.
Thus, the center--surround connectivity is obtained by differentiating the inference objective (see also Supplementary S3 and Supplementary Fig.~\ref{fig:kernel_precision}).

Updating mean activity alone would leave the represented uncertainty unchanged. Differentiating the same variational objective with respect to covariance gives the additional update, $\tau\dot{\mathbf{S}}=-(\mathbf{W}-\beta\mathbf{S}^{-1})$ 
(Supplementary Eq.~\ref{eq:circuit_S}). 
The updated covariance determines uncertainty and dependencies among the expected memory strengths at retrieval. $\mathbf{S}$ therefore can be interpreted as fast weights, as in architectures that distinguish slow and fast connections~\cite{hinton1987using,schmidhuber1992learning,schlag2021linear}. The circuit thus holds a memory in two ways: mean activity, sustained during the delay by the rehearsal term, represents expected memory strength, and the covariance, interpretable as fast synaptic weights, represents uncertainty.

The circuit derivation fixes inducing locations and kernel parameters, whereas MemGP still updates them during maintenance. To assess how this difference affects recall, we compared the recurrent circuit with MemGP on matched trials. For each matched trial, the recurrent circuit and MemGP began from the same posterior obtained after encoding. During maintenance, the recurrent circuit fixed the inducing locations, kernel, and prior mean at their encoding-end values, while MemGP continued to update them. Mean absolute recall error, estimated from 100 reports per trial, agreed closely between the circuit and MemGP (Pearson $r=0.987$, 95\% CI $[0.983,0.990]$; 200 trials pooled across neutral and retrocue conditions; Fig.~\ref{fig:circuit}c).

We next measured how the represented color changed during circuit maintenance. At each epoch, we reconstructed the posterior mean over color at the remembered location and tracked the position of its peak. The peak moved farther from its delay-onset position as maintenance progressed (Spearman $\rho=0.56$, trial-level bootstrap 95\% CI $[0.50,0.62]$; mean displacement $0.02^\circ\to1.26^\circ$ over 100 epochs; 100 trials; Fig.~\ref{fig:circuit}d). The displacement resembles the delay-dependent drift associated with recall errors in prefrontal recordings~\cite{wimmerBumpAttractorDynamics2014}, although the present simulation does not fit those recordings or specify the same noise mechanism.

To test whether uncertainty in the circuit's reports predicts error, we generated one report per item and estimated report variability from separate samples. The circular SD of 200 additional sampled reports was associated with absolute recall error (Spearman $\rho=0.39$, bootstrap 95\% CI $[0.30,0.46]$, $p<0.001$; 100 trials at $N=4$; Fig.~\ref{fig:circuit}e). This association between report variability and error is consistent with evidence that neural uncertainty estimates relate to behavior in perception and working memory~\cite{vanbergenSensoryUncertaintyDecoded2015,liJointRepresentationWorking2021}. 

To test whether the circuit also reproduces the MemGP's selective reduction in uncertainty after a retrocue, we compared posterior-mean and posterior-variance trajectories in cued and neutral delays starting from the same encoded posterior. The conditions produced similar mean trajectories but different variance trajectories: cueing reduced variance at the probed item while neutral maintenance increased it (Fig.~\ref{fig:circuit}f--g). The circuit therefore preserves the distinction observed in MemGP: retrospective attention can reduce uncertainty about memory strength without a corresponding increase in expected strength.

\section{Discussion}

MemGP offers a unified account of working memory across the computational, algorithmic and circuit levels. At the computational level, a posterior over a memory-strength function stores content and uncertainty jointly; at the algorithmic level, a finite set of inducing points is the shared resource that approximates this posterior, and encoding, maintenance, and retrieval are operations (probabilistic transformations) on it. In the simulations, MemGP reproduced load-dependent errors, retrospective cue benefits, and inter-item attraction and repulsion, and each effect could be traced to a specific part of the model. Precision fell with load because all items share the inducing points; a retrocue reduced the variance of the cued memory more than that of the uncued memories, whereas the change in its mean did not differ significantly from theirs; and the direction of inter-item bias followed from how the sampled posterior was normalized. A recurrent circuit derived analytically from MemGP, without training, has the center--surround connectivity assumed by attractor models and agreed closely with MemGP in recall error. Its dynamics also showed delay-dependent drift of the represented color, an association between report variability and recall error, and a selective reduction in uncertainty after a retrocue.

\subsection{Discrete capacity and continuous precision}

MemGP reconciles slot and resource accounts of capacity. 
The number of inducing points is finite, while their locations and the mean and covariance of their function values vary continuously. All remembered items share these inducing points. Increasing the number of items therefore requires the same finite representation to describe more location--color combinations, reducing recall precision in the simulations. A finite number of inducing points does not imply a fixed number of stored objects: an item can depend on several inducing points, and an inducing point can contribute to several items. The model consequently combines a discrete computational resource with continuously varying precision for individual memories. Consistent with this distinction, the variable-precision model described MemGP-generated errors better than the item-limit and slots-plus-averaging models, so a finite number of inducing points did not produce the error distributions expected from a fixed item limit.

Previous models have linked discrete capacity and continuous precision through stochastic sampling, psychophysical similarity, and resource-rational allocation~\cite{schneegansStochasticSamplingProvides2020,schurginPsychophysicalScalingReveals2020,vandenbergResourcerationalTheorySet2018}. These models mainly describe either how a limited resource is allocated among items (encoding) or how a report is generated from memory signals (recall). By contrast, MemGP uses one posterior throughout a trial: the posterior is updated from sensory evidence during encoding, selectively maintained under attention, and sampled to generate a color report.

Among these models, the target confusability competition (TCC) model~\cite{schurginPsychophysicalScalingReveals2020} is most directly comparable to MemGP at retrieval: both select the largest of correlated noisy strength signals for candidate colors. The models differ in the source of these correlations. In MemGP, they arise from the GP posterior, whose covariance depends on the kernel, and the sampled signals undergo spatial divisive normalization before selection. In the TCC model, they are constrained by psychophysical similarity measurements. Using measured perceptual similarity as the GP kernel would let MemGP incorporate the similarity structure of the TCC model. Whether MemGP with this kernel reproduces the error distributions predicted by the TCC model remains to be tested.

\subsection{Attention during maintenance}

The strengthening and protection-from-decay accounts of retrocue benefits~\cite{souzaRetrocuesRetroactiveAttentional2016} both describe the effect of attention on the strength of the memory. In MemGP, attention reweights a maintenance objective that depends on both the posterior mean and covariance, so a cue can affect expected memory strength and uncertainty differently. In the retrocue simulations, the posterior mean of the cued item declined over the delay, and its decline did not differ significantly from that of the uncued items, whereas its posterior variance decreased more. The recurrent circuit showed the same dissociation between cued and neutral delays (Fig.~\ref{fig:circuit}f--g). MemGP therefore improves recall of the cued item mainly by reducing its uncertainty, without strengthening it or significantly slowing its decline. This result relates MemGP to accounts of attention as the control of precision~\cite{feldmanAttentionUncertaintyFreeenergy2010,parrWorkingMemoryAttention2017a} and to behavioral and neural evidence that attentional priority determines the precision of working-memory representations~\cite{baysDynamicShiftsLimited2008,emrichAttentionMediatesFlexible2017,liNeuralMechanismsResource2025}. Experiments combining recall reports, confidence judgments and neural measurements could test whether the behavioral benefit is accompanied by the predicted reduction in uncertainty~\cite{liJointRepresentationWorking2021}.

\subsection{Connectivity and units of the recurrent circuit}

Recurrent models of persistent activity~\cite{wang2001synaptic,wimmerBumpAttractorDynamics2014} assume connectivity with local excitation and surrounding inhibition, and this center--surround pattern maintains a localized bump of activity during the delay~\cite{amariDynamicsPatternFormation1977,compteSynapticMechanismsNetwork2000}. In MemGP, the same pattern analytically follows from the prior. The GP kernel specifies how memory strengths covary across feature combinations under the prior. In the circuit derivation, this prior covariance determines recurrent interactions through its inverse, the prior precision matrix. For the periodic kernel used here, these interactions excite the nearest units and inhibit a surrounding ring (Fig.~\ref{fig:circuit}b). The pattern depends on the smoothness of the kernel and on the conditioning of $\mathbf{K}_{ZZ}$ (Supplementary S3 and Supplementary Fig.~\ref{fig:kernel_precision}).

The units of the derived circuit have a candidate neural counterpart. A neuron with a memory field shows sustained delay activity when the remembered location lies within a limited region of space, and these regions differ across neurons so that together they cover the visual field~\cite{funahashiMnemonicCodingVisual1989}. In MemGP, each inducing point could correspond to the neural population whose memory fields share a region of feature space. The activity of the unit would then be the mean firing rate of this population, representing expected memory strength for that region; the rehearsal term sustains this activity during the delay, and the prior-precision term couples populations tuned to neighboring regions. Because the inducing points are defined over features (e.g., location and color) jointly, the closest counterparts are neurons with conjunctive tuning~\cite{raoIntegrationWhatWhere1997}.

\subsection{Persistent activity, fast weights and activity-silent memory}

The derived circuit holds a memory in both persistent activity and a rapidly updated state, and it assigns them different components of the posterior. The mean activity represents expected memory strength, and the covariance $\mathbf{S}$ represents uncertainty and dependencies among inducing values. The covariance determines the variability and correlations of sampled strengths, which in turn affect the color selected at retrieval. The circuit also contains two kinds of coupling that change on different timescales. The prior-dependent part of the recurrent operator is fixed during the delay, 
whereas $\mathbf{S}$ is updated at every epoch. These two quantities are akin to the slow and fast weight architectures proposed in machine learning~\cite{hinton1987using,schmidhuber1992learning}. Representing expected strength in activity and uncertainty in a rapidly updated state therefore relates the circuit to both persistent-activity accounts and theories in which short-term synaptic changes contribute to memory~\cite{mongilloSynapticTheoryWorking2008,barakPersistentActivityNeural2007}. 

Activity-silent theories further propose that information can remain accessible when stimulus-specific firing is weak or absent~\cite{stokesActivitysilentWorkingMemory2015a,wolffDynamicHiddenStates2017}. The present simulations do not demonstrate storage and recovery of item identity during complete activity silence. The covariance describes uncertainty around the mean; storing the covariance alone need not retain the mean that identifies the remembered items. Demonstrating activity-silent retention would therefore require a mechanism that stores and restores item-specific information when activity ceases and resumes. The model instead motivates tests of how activity and fast synaptic variables jointly determine content and uncertainty, consistent with evidence that persistent and activity-silent processes can interact~\cite{masseCircuitMechanismsMaintenance2019,barbosaInterplayPersistentActivity2020}.

\subsection{Variational inference under computational constraints}

In MemGP, variational inference gives encoding and maintenance the same form of optimization: preserve information while penalizing departure from a prior. During encoding, the likelihood term rewards agreement with sensory observations. During maintenance, it rewards agreement with memory strengths predicted at the preceding epoch. In both stages, the KL term penalizes departure of the inducing-value posterior from the GP prior, and the number of inducing points limits the approximation used to represent that posterior. Attention increases the maintenance likelihood weights near a cued location, giving the corresponding rehearsal targets greater influence on the update. This explicit trade-off between fidelity and regularization relates MemGP to resource-rational theories of memory~\cite{vandenbergResourcerationalTheorySet2018}.

The KL penalty also relates MemGP to efforts to derive neural computation from resource and energy constraints~\cite{gershman2017complex,vafaii2026brain}. In MemGP, this penalty is an information-theoretic cost on the Gaussian posterior over inducing values and does not directly measure metabolic expenditure. In Poisson variational models, by contrast, the KL term becomes proportional to the prior firing rates of model neurons, which gives it an interpretation as metabolic cost~\cite{vafaiiMetabolicCostInformation2026}. A comparable interpretation of MemGP's Gaussian KL penalty would require a biophysical model of how neurons encode and update the posterior mean and covariance.

\subsection{Future directions}

The Gaussian-process formulation separates assumptions about baseline memory strength, feature similarity and computational capacity. The prior mean specifies baseline expected strength. The kernel specifies how strengths at different feature combinations covary and therefore how evidence generalizes across those combinations. The inducing points determine the finite approximation used to represent the posterior. Each component can be changed while retaining the variational inference procedure. Although GP kernels can be defined over other feature spaces, the present behavioral tests are restricted to location and color. Extending to multi-feature binding will require higher-dimensional GPs or structured kernels, in which case deep kernel learning \cite{gortlerVisualExplorationGaussian2019} or spectral mixture kernels could allow the covariance structure to be learned from data.

The present simulations establish qualitative agreement with the tested human behavioral effects, but quantitative fits to individual participants remain to be evaluated. Fitting participant data, potentially with Bayesian optimization~\cite{snoek2012practical}, would test whether distinct parameter settings can be distinguished from their behavioral predictions and whether MemGP predicts recall more accurately than alternative models. In a similar vein, in MemGP time is measured in optimization epochs rather than seconds; relating the dynamics to real delay intervals is important future work.

Finally, in addition to its links to perception through inference from noisy sensory observations, and to decision making through sampling from the posterior at retrieval, MemGP could be linked to long-term memory through its GP prior. Learning across trials could shape the prior mean and kernel to reflect familiar feature combinations and their relationships, consistent with evidence that learned stimulus statistics alter working-memory dynamics~\cite{panichelloErrorCorrectingDynamics2019}. These changes would affect both posterior inference and the recurrent connectivity derived from the prior precision. As the posterior relaxes toward a learned, structured prior during maintenance, recall could become biased toward familiar stimuli. Retaining part of the memory state between trials could also allow the model to test serial dependence, in which a current report depends on previously presented items~\cite{barbosaInterplayPersistentActivity2020}.

\section{Methods}

Full mathematical details are provided in the Supplementary Methods. Here we summarize the key components of MemGP.

\subsection{Model specification}

MemGP implements a sparse variational Gaussian process \cite{titsiasVariationalLearningInducing2009} using GPyTorch \cite{gardnerGPyTorchBlackboxMatrixMatrix2018}. For the continuous-reproduction task in the current study, the input space is two-dimensional: each input is a pair $\bx = (\ell, c)$, with spatial location $\ell \in [-180^\circ, 180^\circ)$ and color $c \in [-180^\circ, 180^\circ)$, both circular with period $360^\circ$. The GP prior mean $m(\bx)$ is constant across feature space, adjusted during inference. The prior covariance is a scaled product of two periodic kernels (Eq.~\ref{eq:prior}). The variational posterior over inducing values is parameterized by a mean $\bm$ and a covariance $\mathbf{S}$, represented through its Cholesky factor~\cite{hensmanScalableVariationalGaussian2015}, at $M = m^2$ inducing points $\bZ = \{(\ell^z_m, c^z_m)\}_{m=1}^M$ initialized on a regular grid (default $m = 10$, $M = 100$), with inducing values $\bu=f(\bZ)$. The inducing positions are inferred and updated during encoding and maintenance. A Gaussian likelihood $p(y_i \mid f_i)$ with fixed, small observation noise completes the generative model (Supplementary Table~\ref{tab:params}).

During cued maintenance, inducing point $m$ receives a spatial attention weight that depends only on its location coordinate. For a cue at $\check{\ell}_j$,
\begin{equation}
    a(\ell^z_m) = 1 + (G - 1)\exp\!\left(-\frac{d(\ell^z_m, \check{\ell}_j)^2}{2\sigma_s^2}\right)
    \label{eq:attention}
\end{equation}
with peak gain $G$, Gaussian spatial width $\sigma_s$, and $d(\cdot,\cdot)$ circular distance. When no cue is presented, all weights equal~$1$ (uniform attention).

\subsection{Simulation protocol}

\textbf{Stimulus generation.} Unless otherwise specified, each trial used $N$ locations sampled without replacement from eight equally spaced positions on a circle, with a common random rotation. Colors were sampled uniformly from $[-180^\circ,180^\circ)$ and converted to integer degrees. A surrogate perceptual procedure generated $n_{\mathrm{obs}}=1000$ observations per item. The location and color of each observation were drawn from Gaussian distributions centered on the item's location and color ($\sigma_\ell=\sigma_c=20^\circ$) and wrapped onto the circular axes. Each observation $(\ell_i, c_i)$ of an item $(\check{\ell}, \check{c})$ was assigned the evidence-strength target $y_i$, based on Gaussian functions of its circular distances from the item (Eq.~\ref{eq:evidence}). These values entered the Gaussian likelihood as regression targets. The model does not simulate the perceptual computations that would generate this evidence.

\textbf{Inference.} Encoding minimizes the negative ELBO (Eq.~\ref{eq:elbo_enc}) using the Adam gradient-based optimizer. Maintenance minimizes the rehearsal objective (Eq.~\ref{eq:elbo_maint}), whose target at epoch $t$ is the posterior means at the inducing points from epoch $t-1$. Both stages update the inducing locations, kernel hyperparameters, prior mean, variational mean, and variational covariance. During cued maintenance, attention reweights the likelihood terms according to Eq.~\ref{eq:attention}. The maintenance optimizer inherits its first- and second-moment estimates from encoding and switches to the maintenance learning rate. Epoch counts, learning rates, and the KL weight $\beta$ are listed in Supplementary Table~\ref{tab:params}. Kernel lengthscales in that table specify their initial values.

\textbf{Retrieval.} One joint sample was drawn from the latent GP posterior over 360 evenly spaced color values at every occupied item location, preserving covariance across both colors and items. Subtracting the minimum sampled strength across all locations and colors from every sampled strength made the responses nonnegative. Spatial divisive normalization~\cite{carandiniSummationDivisionNeurons1994,carandiniNormalizationCanonicalNeural2012} was then applied, and the color with the maximum normalized response was reported (Supplementary S1.5). The parameters of spatial divisive normalization were consistent across the set-size, retrocue, and bias experiments. Three analyses used repeated draws from the same posterior: swap rates (100 reports per probe; Fig.~\ref{fig:setsize}m), matched-trial recall error of the recurrent circuit (100 reports per trial; Fig.~\ref{fig:circuit}c), and report uncertainty (200 additional reports per item; Fig.~\ref{fig:circuit}e). All other analyses used one draw per trial.

\subsection{Experiments}

\textbf{Set size.} Load was varied over $N \in \{1, 2, 4, 6, 8\}$ (20 simulated subjects, 100 trials per set size; uniform attention). As in human continuous-report tasks, one randomly located item was probed per trial, so every set size gave each subject the same number of errors. Circular SD and precision (inverse circular SD) were computed per subject and summarized as the subject mean $\pm$ s.e.m. For non-target-relative errors at $N=4$, the probed report on each trial was compared with the color of each of the three non-target items, and the signed circular differences (report minus non-target color) were pooled across trials and subjects (Fig.~\ref{fig:setsize}i). Hyperparameter sweeps repeated the set-size series while varying inducing-grid size $M \in \{25, 49, 100, 196\}$, KL weight $\beta \in \{0, 1, 5, 20\}$, or initial kernel lengthscale $\lambda \in \{5, 10, 20, 40\}$ (location and color kernels set equal).

\textbf{Misbinding.} Swap rates as a joint function of set size and target--distractor spatial distance used $N \in \{2, 4, 6, 8\}$ (20 simulated subjects, 30 displays per set size). Every item in each display was probed, with 100 reports generated from independent posterior samples for each probe. A report counted as a swap toward a given non-target if it fell within $30^\circ$ of that item's color and outside $30^\circ$ of the target. Each probed item and each non-target in the same display formed a target--distractor pair. Because items occupied positions spaced $45^\circ$ apart, the circular spatial distance between the two items in a pair took one of four values, $\{45^\circ, 90^\circ, 135^\circ, 180^\circ\}$. Swap rates were averaged within each combination of set size and distance (Fig.~\ref{fig:setsize}m).

\textbf{Retrocue.} Neutral trials used uniform attention throughout a 100-epoch delay. Cued trials applied Eq.~\ref{eq:attention} from epoch 50, with attention weights centered on the probed item's location. Load dependence compared $N=2$ and $N=4$ (20 simulated subjects, 100 trials per condition). Cue-flexibility trials ($N=4$, 200-epoch delay; 20 simulated subjects, 100 trials; Fig.~\ref{fig:retrocue}h) applied the same gain first to a randomly chosen item A (epochs 50--99) and then to a different item B (epochs 100--199). Absolute recall error for A and B was read out at the end of each cue window.

\textbf{Attraction/repulsion bias.} Displays used a fixed target at location $0^\circ$. At $N=2$, one distractor sat at $90^\circ$. At $N=6$, the six items occupied positions spaced $60^\circ$ apart. The five distractors lay on the same side of the target in color, at color distances $D-s$, $D-s/2$, $D$, $D+s/2$ and $D+s$ from it, where $D$ is the nominal color distance and $s=\min(8^\circ, \max(2^\circ, 0.2D))$. The nominal color distance $D$ was swept over $\{20, 45, 75, 105, 135\}^\circ$ (20 simulated subjects, 100 trials per condition). Cued vs.\ neutral comparisons and the encoding-noise comparison ($\sigma_c=10^\circ$ vs.\ $60^\circ$) used $N=2$ at $135^\circ$. For the precision--repulsion correlation (Fig.~\ref{fig:bias}g), each of 24 simulated subjects drew encoding noise from Uniform$[10^\circ,40^\circ]$ and was tested at color distances $\{20, 40, 60, 80, 100, 120, 140\}^\circ$; circular SD and the mean signed error (degrees; trials with absolute error above $40^\circ$ excluded) were averaged across distances. Human subject means in Fig.~\ref{fig:bias}f were recomputed from \citet{chunharasAdaptivePerspectiveVisual2022a} Exp.~4, aggregating over encoding $\times$ distance cells. Bias in Fig.~\ref{fig:bias}b--e was the model-free measure in Eq.~\ref{eq:bias}.

\textbf{Recurrent circuit implementation.}
Matched-trial recall compared MemGP to the recurrent circuit on 100 displays ($N=4$). Each display was encoded once and then maintained under both the neutral and the cued condition, giving 200 data points that were pooled in Fig.~\ref{fig:circuit}c. For each trial, MemGP and the recurrent circuit begin from the same posterior obtained after encoding. The recurrent circuit fixes the inducing locations, kernel and prior mean at encoding end, while MemGP continues to update them through the delay. The comparison in Fig.~\ref{fig:circuit}c uses mean absolute error over 100 sampled reports per trial, with matched random seeds for readout. The recurrent circuit retains the encoding optimizer state and updates the variational mean and Cholesky covariance in GPyTorch's whitened coordinates~\cite{hensmanMCMCVariationallySparse2015,matthewsScalableGaussianProcess2016}. Adam updates depend on the choice of coordinates, so the circuit and MemGP share one numerical update rule and differ only in which parameters are free during the delay.

Bump displacement in the recurrent circuit (Fig.~\ref{fig:circuit}d) is the movement of the peak read out from $\bm$: the argmax, on a $0.1^\circ$ color axis, of the posterior mean at the item's location, measured from its position at delay onset ($N=1$, neutral maintenance, $100$ trials). The argmax was taken within $45^\circ$ of the peak's position at the previous epoch, starting from the item's color, so that a second peak elsewhere on the color axis is not read out as displacement. On the inducing grid, that mean equals $\bm$; off the grid, it is the kernel interpolation of $\bm$ (Supplementary~S3.2). 

For Fig.~\ref{fig:circuit}e ($N=4$, neutral maintenance, $100$ trials), each of the four items on each trial received one sampled report from the recurrent circuit's posterior defined by the circuit's mean and covariance at the end of the delay; uncertainty was the circular SD of 200 additional reports from the same posterior and readout. The circular SD and the independent report error therefore used separate draws.

The mean and variance trajectories (Fig.~\ref{fig:circuit}f,g) were obtained by running cued and neutral maintenance in the recurrent circuit from the same encoded posterior ($N=4$, $100$ trials), with the cue applied from epoch 50, and are plotted as the change from each trial's value at cue onset.

\subsection{Model comparison}

The comparison fitted the benchmark models of \citet{vandenbergVariabilityEncodingPrecision2012} to MemGP's simulated errors to assess which descriptive model best summarized those errors. The fitted models consist of Item-Limit (IL), Equal-Precision (EP), Slots-plus-Averaging (SA), and Variable-Precision (VP), which were fitted separately to each simulated subject's signed errors ($N \in \{1, 2, 4, 6, 8\}$, 500 trials per subject) by maximum likelihood with differential evolution (following \citet{vandenbergVariabilityEncodingPrecision2012}). Akaike information criterion (AIC), Bayesian information criterion (BIC), and the small-sample correction to AIC (AICc) were computed per subject (Supplementary Table~\ref{tab:ic}). 

In addition to comparing information criteria, we assessed whether each fitted model reproduced the mixture statistics of MemGP-generated errors. This comparison (Supplementary Fig.~\ref{fig:model_comparison}) followed \citet{vandenbergVariabilityEncodingPrecision2012}: for each subject and model, 10 synthetic datasets of the subject's size were generated from the maximum-likelihood parameters, the uniform--von~Mises mixture was fit to each, and the estimated von~Mises mixture weight $w$ and the circular SD of that component were averaged to give that subject's prediction. Ribbons are $\pm$ s.e.m.\ of the predictions across subjects, and root mean squared error (RMSE) between predicted mixture parameters and those estimated from MemGP-generated errors was computed over all subjects and set sizes.

\subsection{Statistical analysis}

Error bars on behavioral simulation panels are s.e.m.\ across 20 simulated subjects that differ in random seed, with a few exceptions: maintenance traces (Fig.~\ref{fig:retrocue}k--l) instead show means $\pm$ s.e.m.\ across 100 trials; the precision--repulsion analysis (Fig.~\ref{fig:bias}g) has 24 simulated subjects (matching the human experiment), each with its own encoding noise level; circuit trajectory bands are bootstrap 95\% intervals as specified in Fig.~\ref{fig:circuit}.

Retrocue benefits (Fig.~\ref{fig:retrocue}f) were paired $t$-tests on subject-mean absolute error (20 subjects, 100 trials per condition; cued vs.\ neutral). The load comparison was an independent-samples $t$-test on those subject-level benefits (one-tailed, $H_1$: greater benefit at $N\!=\!4$ than at $N\!=\!2$). The cue-reallocation test (Fig.~\ref{fig:retrocue}h) was a paired $t$-test, across 20 subjects, of the trade-off $\Delta\mathrm{A}-\Delta\mathrm{B}$ (error change for the abandoned item minus that for the newly cued item), computed from each subject's mean errors over 100 trials. The maintenance-dynamics comparisons (Fig.~\ref{fig:retrocue}k--l) were paired $t$-tests of the end-of-delay change from cue onset, cued versus the mean of the uncued items, across 100 trials: predictive mean in (k) and posterior variance in (l).

Bias used Eq.~\ref{eq:bias} with the simulated subject as the unit ($n = 20$, 100 trials per condition). Precision--repulsion correlations (Fig.~\ref{fig:bias}f--g) are Pearson $r$ on subject means. Other correlations are Pearson $r$ or Spearman $\rho$ as labeled at each use, with 95\% confidence intervals (Fisher $z$ for Pearson; bootstrap for Spearman where noted). The uncertainty--error correlation in Fig.~\ref{fig:circuit}e uses 400 item-level observations from 100 displays; its 95\% CI was obtained by resampling displays, so that the four items from each display were kept together.

\section*{Data and code availability}

The code for the MemGP model, the simulations, the analyses, and the figures, together with the simulation outputs that the figures use, is available at \url{https://github.com/Daniel-Gong/MemGP-Code}. This study collected no new human data. The human data replotted in Figs.~\ref{fig:setsize}--\ref{fig:bias} come from published studies~\cite{baysPrecisionVisualWorking2009b,souzaFocusedAttentionImproves2014,vanMoorselaarForgottenButNot2015,chunharasAdaptivePerspectiveVisual2022a}; copies of the files used, with their sources, are provided in the \texttt{data} folder of the same repository.

\section*{Author contributions}

Conceptualization: D.G., M.B.\ and I.Y. Code: D.G. Formal analysis: D.G. Investigation: D.G. Visualization: D.G. Writing (original draft): D.G.\ and I.Y. Writing (review \& editing): D.G., M.B.\ and I.Y. Supervision: I.Y.

\section*{Competing interests}

The authors declare no competing interests.

\bibliographystyle{unsrtnat}
\IfFileExists{references.bib}{\bibliography{references}}{\bibliography{paper/references}}

\clearpage
\appendix
\setcounter{figure}{0}
\setcounter{table}{0}
\renewcommand{\thefigure}{S\arabic{figure}}
\renewcommand{\thetable}{S\arabic{table}}
\setcounter{equation}{0}
\renewcommand{\theequation}{S\arabic{equation}}

\section*{Supplementary Materials}

\subsection*{S1. Mathematical formulation}

\subsubsection*{S1.1 Gaussian process prior}

\begin{figure}[htbp]
    \centering
    \includegraphics[width=\textwidth,height=0.58\textheight,keepaspectratio]{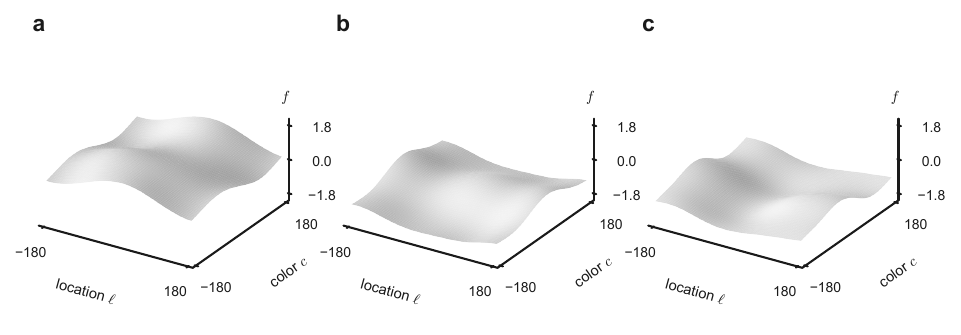}
    \caption{\textbf{The GP prior is a distribution over smooth random functions on the circular location--color torus.}
    \textbf{(a--c)} Independent zero-mean draws $f\sim\GP(0,k)$, each a map $[-180,180)^2\to\R$.}
    \label{fig:prior}
\end{figure}

The latent memory-strength function $f: \mathcal{X} \to \R$ assigns a strength to each location--color combination, where $\mathcal{X} = [-180, 180)^2$ is the circular location--color space and each input is a pair $\bx = (\ell, c)$. The GP prior (Eq.~\ref{eq:prior}) is
\[
f \sim \GP\bigl(m(\bx), \, k(\bx, \bx')\bigr)
\]
where $m(\bx)$ is the prior mean, constant across feature space, initialized at 0 and adjusted during inference. The prior covariance between two location--color combinations is
\begin{equation}
k(\bx, \bx') = \sigma^2_f \cdot k_{\mathrm{per}}(\ell, \ell'; \lambda_\ell, P) \cdot k_{\mathrm{per}}(c, c'; \lambda_c, P)
\label{eq:kernel}
\end{equation}
The covariance is the product of a location kernel and a color kernel, multiplied by the output scale $\sigma_f^2$. Each periodic kernel is
\begin{equation}
k_{\mathrm{per}}(x, x'; \lambda, P) = \exp\!\left(-\frac{2 \sin^2(\pi |x - x'| / P)}{\lambda}\right)
\label{eq:kper}
\end{equation}
with period $P = 360^\circ$ for both dimensions. The denominator follows the GPyTorch PeriodicKernel convention: it contains $\lambda$, not its square. The parameter $\lambda$ determines how covariance changes with distance along the circular dimension. The output scale $\sigma^2_f$ is updated as part of a \texttt{ScaleKernel} wrapper. Figure~\ref{fig:prior} shows independent draws of $f$ from $\GP(0,k)$.

\subsubsection*{S1.2 Sparse variational approximation}

An observation consists of a location--color combination $\bx_i$ and an evidence strength $y_i$. The Gaussian likelihood $p(y_i \mid f_i)$ relates the observed strength to the latent function value $f_i = f(\bx_i)$. The exact posterior at a query location $\bx^*$ is $p(f^* \mid \mathbf{X}, \mathbf{y}, \bx^*)$, $f^* = f(\bx^*)$. We approximate it using $M$ inducing points $\bZ = \{(\ell^z_m, c^z_m)\}_{m=1}^M$, $\bZ \in \R^{M \times 2}$. The inducing values are $\bu = f(\bZ)$. The variational mean $\bm \in \R^M$ and covariance $\mathbf{S} \in \R^{M \times M}$ describe the distribution of these inducing values. The diagonal entries of $\mathbf{S}$ are marginal variances; off-diagonal entries are covariances between inducing values. The implementation parameterizes $\mathbf{S}$ through a Cholesky factor, whose product with its transpose gives the covariance (\texttt{CholeskyVariationalDistribution}). The variational distribution (Eq.~\ref{eq:qu}) is
\[
q(\bu) = \mathcal{N}(\bu; \, \bm, \, \mathbf{S}).
\]
To predict strength at a queried location--color combination, the model integrates the GP conditional over the distribution of inducing values. This gives the approximate posterior predictive distribution:
\begin{equation}
q(f(\bx^*)) = \mathcal{N}\!\left(
    \mu_0 + \mathbf{k}_*^\top \mathbf{K}_{ZZ}^{-1}(\bm-\mu_0\mathbf{1}), \;
    k_{**} - \mathbf{k}_*^\top \mathbf{K}_{ZZ}^{-1}(\mathbf{K}_{ZZ} - \mathbf{S})\mathbf{K}_{ZZ}^{-1} \mathbf{k}_*
\right)
\label{posterior-predictive}
\end{equation}
Here $\mu_0$ is the constant prior mean. The vector $\mathbf{k}_* = k(\bZ, \bx^*)$ contains prior covariances between the queried combination and the inducing points; $k_{**} = k(\bx^*, \bx^*)$ is the prior variance at the query. The matrix $\mathbf{K}_{ZZ} = k(\bZ, \bZ)$ contains prior covariances among the inducing values. The first argument of the Gaussian is the predicted mean strength, and the second is its variance.

Inducing points are initialized on a regular $m \times m$ grid ($m = 10$ by default, $M = 100$) with centers offset by half the bin width to ensure uniform coverage. Their positions are inferred during optimization (\texttt{learn\_inducing\_locations = True}).

\subsubsection*{S1.3 Encoding objective}

Given $N$ items $\{(\check{\ell}_i, \check{c}_i)\}_{i=1}^N$, the surrogate perceptual procedure generates $n_{\mathrm{obs}}$ observations per item $\bx_{ij} = (\check{\ell}_i + \epsilon_{\ell}, \check{c}_i + \epsilon_c)$ with $\epsilon_\ell \sim \mathcal{N}(0, \sigma_\ell^2)$, $\epsilon_c \sim \mathcal{N}(0, \sigma_c^2)$, wrapped to $[-180, 180)$. The evidence-strength targets $y_{ij}=w_{ij}$ are
\begin{equation}
w_{ij} = \exp\!\left(-\frac{d(\ell_{ij}, \check{\ell}_i)^2}{2\sigma_\ell^2}\right) \cdot \exp\!\left(-\frac{d(c_{ij}, \check{c}_i)^2}{2\sigma_c^2}\right),
\label{eq:evidence}
\end{equation}
where $d(\cdot,\cdot)$ is circular distance.

The encoding ELBO is maximized using Adam with learning rate $\eta_{\mathrm{enc}} = 0.05$ for $T_{\mathrm{enc}} = 100$ epochs. The Gaussian likelihood noise variance is fixed (not inferred; Table~\ref{tab:params}).

\subsubsection*{S1.4 Maintenance objective}

During maintenance, the model uses its preceding predictions as observations in the variational objective. The likelihood is evaluated at the current inducing locations $\bZ^{(t)}=\{\bz_m^{(t)}\}_{m=1}^M$. Self-rehearsal targets are initialized at the end of encoding as the posterior mean of inducing values and then updated after each epoch from the current posterior mean,
\begin{equation}
\tilde{y}_m^{(0)} = \E_{q(f)}[f(\bz_m)]\big|_{t=T_{\mathrm{enc}}},\qquad
\tilde{y}_m^{(t)} = \E_{q(f)}[f(\bz_m^{(t)})]\big|_{t-1}.
\label{eq:rehearsal}
\end{equation}

The maintenance objective (Eq.~\ref{eq:elbo_maint}) is optimized with Adam (learning rates and epoch counts in Table~\ref{tab:params}). The KL coefficient $\beta$ controls regularization toward the prior and is set to $\beta=2$ in the baseline configurations. The effect of KL regularization on posterior variance depends on the rehearsal term, the attention weights and the learned prior; it does not require variance to increase at every point on every update. Before cue onset (epoch $t_{\mathrm{cue}}$), attention weights are uniform ($a(\ell^z_m) = 1$). After cue onset, the spatial gain function (Eq.~\ref{eq:attention}) with parameters $G$ (attended gain) and $\sigma_s$ (spatial width) is applied; because the inducing locations move during the delay, the gains are recomputed at the current locations in every epoch.

\subsubsection*{S1.5 Retrieval}

A single joint sample $f \sim \mathcal{N}(\boldsymbol{\mu}, \boldsymbol{\Sigma})$ is drawn from the GP posterior predictive over 360 evenly spaced color values at every occupied item location, preserving inter-color and inter-item covariance. Let $f_i(c) = f(\check{\ell}_i, c)$ denote the sampled memory-strength profile over color at the location $\check{\ell}_i$ of item $i$, and let $f_t(c) = f(\ell_{\mathrm{probe}}, c)$ denote the profile at the probed location. The probed profile is divisively normalized by a pool of all item profiles from the same sample. To make all responses nonnegative, the same minimum sampled strength is subtracted from every value, $f_i^+(c) = f_i(c) - \min f$ (and likewise $f_t^+$), where the minimum is taken over every item location and color. Unlike rectification, which would set negative strengths to zero, this common shift keeps the color profile of every item in the pool, including items with little evidence. Each item enters the pool with a spatial surround weight
\begin{equation}
w_i = \exp\!\left(-\,d(\check{\ell}_i, \ell_{\mathrm{probe}})^2 \,/\, 2\sigma_{\mathrm{pool}}^2\right).
\label{eq:pool_weight}
\end{equation}
The probed item has zero distance from the probe and therefore has $w=1$. The pool is the weighted sum of shifted profiles,
\begin{equation}
\mathrm{pool}(c) = \textstyle\sum_i w_i\, f_i^+(c).
\label{eq:pool}
\end{equation}
For each candidate color, the probed item's shifted strength is divided by this pool plus a baseline term. The normalized response and the selected report are
\begin{equation}
r(c) = \frac{f_t^+(c)}{\sigma_b \cdot \max_c f_t^+(c) + \mathrm{pool}(c)},
\qquad \hat{c} = \arg\max_c\, r(c),
\label{eq:readout}
\end{equation}
where $\sigma_b$ is the baseline (semi-saturation) fraction. Two parameters govern the readout: the normalization-pool spatial width $\sigma_{\mathrm{pool}}$, which determines how rapidly each item's weight $w_i$ decreases with distance from the probe, and the baseline fraction $\sigma_b$, which scales the maximum target response in the denominator (Table~\ref{tab:params}). The baseline term scales with the maximum shifted response of the probed item rather than being a fixed constant as in the standard normalization equation~\cite{carandiniNormalizationCanonicalNeural2012}; this makes the readout invariant to the overall scale of the sampled strengths, which decline during the delay.

\begin{table}[htbp]
\centering
\caption{\textbf{Baseline settings across behavioral experiments.}}
\label{tab:params}
\small
\begin{tabular}{lccc}
\toprule
\textbf{Parameter} & \textbf{Set-size} & \textbf{Retrocue} & \textbf{Bias} \\
\midrule
$n_{\mathrm{obs}}$ per item & 1000 & 1000 & 1000 \\
$\sigma_\ell$ (loc sampling noise) & $20^\circ$ & $20^\circ$ & $20^\circ$ \\
$\sigma_c$ (color sampling noise) & $20^\circ$ & $20^\circ$ & $20^\circ$ \\
$m$ (inducing grid side) & 10 & 10 & 10 \\
$\lambda_\ell$ (initial loc lengthscale) & 10.0 & 10.0 & 10.0 \\
$\lambda_c$ (initial color lengthscale) & 10.0 & 10.0 & 10.0 \\
$\sigma_n^2$ (likelihood noise) & $10^{-4}$ & $10^{-4}$ & $10^{-4}$ \\
$T_{\mathrm{enc}}$ (encoding epochs) & 100 & 100 & 100 \\
$T_{\mathrm{maint}}$ (maintenance epochs) & 100 & 100$^{\mathrm{a}}$ & 100 \\
$\eta_{\mathrm{enc}}$ (encoding LR) & 0.05 & 0.05 & 0.05 \\
$\eta_{\mathrm{maint}}$ (maintenance LR) & 0.01 & 0.01 & 0.01 \\
$\beta$ (KL weight) & 2 & 2 & 2 \\
$\sigma_s$ (attention spatial width) & -- & $20^\circ$ & -- \\
$G$ (attended gain) & -- & 40.0 & -- \\
$\sigma_{\mathrm{pool}}$ (normalization-pool spatial width) & $60^\circ$ & $60^\circ$ & $60^\circ$ \\
$\sigma_b$ (baseline fraction) & 0.8 & 0.8 & 0.8 \\
\bottomrule
\end{tabular}
\\[0.4em]
{\footnotesize Inducing positions, kernel lengthscales, and prior mean are updated during inference. Bias trials with a target cue use the retrocue attention settings. $^{\mathrm{a}}$The cue-flexibility simulation (Fig.~\ref{fig:retrocue}h) uses $T_{\mathrm{maint}} = 200$ to accommodate the two cue windows (cue A, epochs 50--99; cue B, epochs 100--199).}
\end{table}

\subsection*{S2. Information-theoretic model comparison}

Following \citet{vandenbergVariabilityEncodingPrecision2012}, we compared descriptive models fitted to MemGP errors. For each descriptive model $\mathcal{M}$, we estimated $k$ free parameters by maximum likelihood estimation (MLE) from $n$ recall errors. We compared fits using Akaike information criterion (AIC), Bayesian information criterion (BIC), and the small-sample correction to AIC (AICc):
\begin{align}
    \mathrm{AIC} &= 2k - 2\ln \hat{L} \\
    \mathrm{BIC} &= k \ln n - 2\ln \hat{L} \\
    \mathrm{AICc} &= \mathrm{AIC} + \frac{2k(k+1)}{n - k - 1}
\end{align}
Here $\hat{L}$ is the maximized likelihood. Each criterion combines a term measuring fit with a penalty for the number of fitted parameters. Lower AIC/BIC indicates a preferred model under the corresponding criterion (see Table~\ref{tab:ic}). Parameter counts: IL ($k=3$: $K, \kappa_{\mathrm{mem}}, \kappa_r$), SA ($k=3$: $K, J_1, \kappa_r$), EP ($k=3$: $J_1, \alpha, \kappa_r$), VP ($k=4$: $\bar{J}_1, \alpha, \tau, \kappa_r$).

\begin{table}[htbp]
\centering
\caption{\textbf{Information-theoretic model comparison.}
Mean $\Delta$AIC and $\Delta$BIC ($\pm$ s.e.m.) across 20 simulated subjects. Each model was fit by MLE to that subject's signed errors ($n = 500$: 100 trials $\times$ set sizes 1, 2, 4, 6, 8). For each subject, $\Delta$AIC and $\Delta$BIC are computed relative to the model with the lowest value of that criterion; positive values indicate a worse fit. VP had the lowest AIC and BIC in all 20 simulated subjects. These criteria compare descriptive models fitted to MemGP-generated recall errors.}
\label{tab:ic}
\small
\begin{tabular}{lcrrc}
\toprule
\textbf{Model} & $k$ & $\Delta$\textbf{AIC} & $\Delta$\textbf{BIC} & \textbf{Best (AIC)} \\
\midrule
VP & 4 & $0.0 \pm 0.0$ & $0.0 \pm 0.0$ & 20/20 \\
IL & 3 & $89.0 \pm 3.5$ & $84.8 \pm 3.5$ & 0/20 \\
SA & 3 & $91.7 \pm 7.2$ & $87.5 \pm 7.2$ & 0/20 \\
EP & 3 & $131.8 \pm 8.7$ & $127.6 \pm 8.7$ & 0/20 \\
\bottomrule
\end{tabular}
\end{table}

\begin{figure}[htbp]
    \centering
    \includegraphics[width=\textwidth,height=0.78\textheight,keepaspectratio]{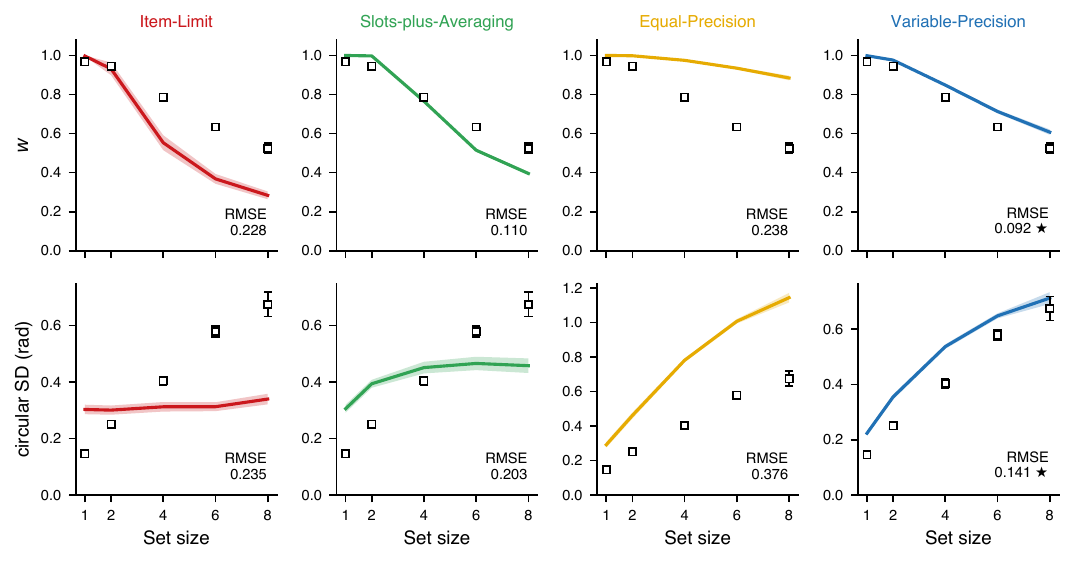}
    \caption{\textbf{Formal model comparison.}
    Uniform--von~Mises mixture statistics extracted from MemGP simulation errors (open squares, mean $\pm$ s.e.m.\ across 20 simulated subjects) compared with predictions from four benchmark models fit by maximum likelihood to each subject's errors ($N \in \{1, 2, 4, 6, 8\}$, 100 trials each). Top row: mixture weight $w$ (probability of non-guess response). Bottom row: circular SD of the von~Mises component. Colored lines show mean model predictions across subjects; shaded ribbons are $\pm$ s.e.m.\ across subjects. RMSE, computed over all subjects and set sizes between each subject's prediction and statistics, quantifies goodness of fit; $\bigstar$ marks the lowest RMSE. VP had the lowest RMSE for both mixture statistics and the lowest AIC and BIC (Supplementary Table~\ref{tab:ic}).}
    \label{fig:model_comparison}
\end{figure}

\subsection*{S3. Recurrent circuit implementation of MemGP}

To derive recurrent equations, we hold the inducing locations, kernel parameters, likelihood noise, and prior mean fixed. Only the variational mean and covariance are updated in the derivation. We express the variational objective in terms of the mean $\bm$ and covariance $\mathbf{S}$ of the inducing values $\bu=f(\bZ)$. The inverse prior covariance, or prior precision, determines interactions between units that remain fixed under these assumptions, while $\mathbf{S}$ is a rapidly updated covariance state. Attention changes diagonal gains in the effective operator $\mathbf{W}$. If $\bm$ is read as mean activity and $\mathbf{S}$ as fast synaptic weights, the resulting equations describe a recurrent circuit implementation of MemGP.

The derivation starts from closed-form expressions for the two terms of the variational objective: the expected log-likelihood and the KL divergence. The likelihood is Gaussian, $p(y\mid f)=\mathcal{N}(y;f,\sigma^2)$, and the prior on the inducing values is $p(\bu)=\mathcal{N}(\mu_0\mathbf{1},\mathbf{K}_{ZZ})$, with $\mu_0$ the constant prior mean. The variational posterior is $q(\bu)=\mathcal{N}(\bm,\mathbf{S})$. Let $q(f_i)=\mathcal{N}(\mu_i,v_i)$ denote the marginal posterior of one latent strength $f_i$, with mean $\mu_i$ and variance $v_i$. Averaging the log-likelihood over this posterior gives
\begin{equation}
\E_q[\log p(y_i\mid f_i)]
=-\frac12\log(2\pi\sigma^2)
-\frac{(y_i-\mu_i)^2+v_i}{2\sigma^2}.
\label{eq:gauss_ell}
\end{equation}

The KL divergence measures the departure of the Gaussian posterior $q(\bu)$ from the Gaussian prior $p(\bu)$. Its closed-form expression is
\begin{equation}
\begin{aligned}
\KL\bigl(q(\bu)\,\|\,p(\bu)\bigr)
&=\frac12(\bm-\mu_0\mathbf{1})^\top\mathbf{K}_{ZZ}^{-1}(\bm-\mu_0\mathbf{1})
+\frac12\mathrm{tr}(\mathbf{K}_{ZZ}^{-1}\mathbf{S})\\
&\quad-\frac12\log|\mathbf{S}|
+\frac12\log|\mathbf{K}_{ZZ}|-\frac{M}{2}.
\end{aligned}
\label{eq:kl_gauss}
\end{equation}

\subsubsection*{S3.1 Encoding}

For $n$ perceptual observations, Eq.~\ref{eq:gauss_ell} gives the negative expected log-likelihood in Eq.~\ref{eq:elbo_enc}:
\begin{equation}
-\sum_{i=1}^{n}\E_q[\log p(y_i\mid f_i)]
=\frac{n}{2}\log(2\pi\sigma^2)
+\frac{1}{2\sigma^2}\lVert\mathbf{y}-\boldsymbol{\mu}\rVert^2
+\frac{1}{2\sigma^2}\sum_{i=1}^{n}v_i.
\label{eq:enc_nll}
\end{equation}

To differentiate the encoding objective with respect to the inducing mean $\bm$, we first determine how the two data terms in Eq.~\ref{eq:enc_nll} depend on $\bm$: the squared residual depends on $\bm$ through the predictive mean $\boldsymbol{\mu}$, and the variance sum $\sum_i v_i$ does not depend on $\bm$. Let $\boldsymbol{\Phi}=\mathbf{K}_{XZ}\mathbf{K}_{ZZ}^{-1}$, where $\mathbf{K}_{XZ}$ contains prior covariances between observation locations and inducing points. Multiplying by $\boldsymbol{\Phi}$ maps inducing-value deviations from the prior mean to predicted deviations at the observation locations. For a constant prior mean, the predictive mean in Eq.~\ref{posterior-predictive} is $\boldsymbol{\mu}=\mu_0\mathbf{1}_n+\boldsymbol{\Phi}(\bm-\mu_0\mathbf{1}_M)$, which is linear in $\bm$. Collecting the terms that do not contain $\bm$ into the adjusted observation vector $\mathbf{y}_c=\mathbf{y}-\mu_0\mathbf{1}_n+\boldsymbol{\Phi}\mu_0\mathbf{1}_M$ gives the residual $\mathbf{y}-\boldsymbol{\mu}=\mathbf{y}_c-\boldsymbol{\Phi}\bm$. The predictive variance at $\bx_i$ is
\begin{equation}
v_i=k(\bx_i,\bx_i)-\mathbf{k}_i^\top\mathbf{K}_{ZZ}^{-1}\mathbf{k}_i
+\mathbf{k}_i^\top\mathbf{K}_{ZZ}^{-1}\mathbf{S}\,\mathbf{K}_{ZZ}^{-1}\mathbf{k}_i,
\label{eq:pred_var}
\end{equation}
where $\mathbf{k}_i$ is the vector of prior covariances between observation $\bx_i$ and the inducing points. This expression contains $\mathbf{S}$ and kernel terms but not $\bm$, so $\sum_i v_i$ is constant with respect to $\bm$.

Define $\bar{\mathcal{L}}_{\mathrm{enc}}=\frac{1}{n}\mathcal{L}_{\mathrm{enc}}$, substitute~\eqref{eq:enc_nll} and~\eqref{eq:kl_gauss} into~\eqref{eq:elbo_enc}, and keep the terms that contain $\bm$:
\begin{align}
\bar{\mathcal{L}}_{\mathrm{enc}}(\bm)
&=\frac{1}{n}\Biggl(
\frac{1}{2\sigma^2}\lVert\mathbf{y}_c-\boldsymbol{\Phi}\bm\rVert^2
+\frac12(\bm-\mu_0\mathbf{1})^\top\mathbf{K}_{ZZ}^{-1}(\bm-\mu_0\mathbf{1})
\Biggr)+\text{const} \nonumber\\
&=\frac{1}{2\sigma^2 n}\lVert\mathbf{y}_c-\boldsymbol{\Phi}\bm\rVert^2
+\frac{1}{2n}(\bm-\mu_0\mathbf{1})^\top\mathbf{K}_{ZZ}^{-1}(\bm-\mu_0\mathbf{1})
+\text{const}.
\label{eq:enc_quad}
\end{align}
where the constant collects everything independent of $\bm$, including $\sum_i v_i$, $\mathrm{tr}(\mathbf{K}_{ZZ}^{-1}\mathbf{S})$, and $\log|\mathbf{S}|$.

Differentiating~\eqref{eq:enc_quad} with respect to $\bm$,
\begin{equation}
\nabla_{\bm}\bar{\mathcal{L}}_{\mathrm{enc}}
=\frac{1}{\sigma^2 n}\boldsymbol{\Phi}^\top(\boldsymbol{\Phi}\bm-\mathbf{y}_c)
+\frac{1}{n}\mathbf{K}_{ZZ}^{-1}(\bm-\mu_0\mathbf{1})
=\mathbf{W}_{\mathrm{enc}}\bm-\mathbf{b}_{\mathrm{enc}},
\label{eq:enc_grad}
\end{equation}
where
\begin{equation}
\mathbf{W}_{\mathrm{enc}}
=\frac{1}{\sigma^2 n}\boldsymbol{\Phi}^\top\boldsymbol{\Phi}
+\frac{1}{n}\mathbf{K}_{ZZ}^{-1},
\qquad
\mathbf{b}_{\mathrm{enc}}
=\frac{1}{\sigma^2 n}\boldsymbol{\Phi}^\top\mathbf{y}_c
+\frac{1}{n}\mathbf{K}_{ZZ}^{-1}\mu_0\mathbf{1}.
\label{eq:enc_Wb}
\end{equation}

Gradient flow on Eq.~\ref{eq:enc_quad} gives $\tau\dot{\bm}=-\mathbf{W}_{\mathrm{enc}}\bm+\mathbf{b}_{\mathrm{enc}}$ in the recurrent circuit. The input drive includes $\boldsymbol{\Phi}^\top\mathbf{y}_c$, which projects the adjusted observations onto the inducing units. The recurrent operator contains a likelihood term determined by $\boldsymbol{\Phi}^\top\boldsymbol{\Phi}$ and a prior term determined by the precision matrix. Because $\boldsymbol{\Phi}^{\top}\boldsymbol{\Phi}$ is positive semidefinite and $\mathbf{K}_{ZZ}^{-1}$ is positive definite, $\mathbf{W}_{\mathrm{enc}}\succ0$, and $\bm$ converges to the unique minimizer of~\eqref{eq:enc_quad} from every initial value. This convergence statement assumes a fixed objective and applies to the ideal gradient flow; finite Adam updates need not reach its equilibrium.

\subsubsection*{S3.2 Maintenance}

During maintenance, the encoding observations are replaced by the network's preceding predictions. The maintenance objective (Eq.~\ref{eq:elbo_maint}) therefore weights the expected log-likelihood of the network's own previous posterior means by the attention gains, and multiplies the KL by $\beta$. The rehearsal targets $\mathbf{t}=\tilde{\mathbf{y}}$ are the posterior mean strengths from the preceding epoch. Each epoch holds these targets fixed while updating the variational mean and covariance, then replaces the targets with the updated predicted means. The maintenance likelihood is evaluated at the inducing points, where the predictive moments are $\bm$ and $\mathbf{S}$. Let $\mathbf{A}=\mathrm{diag}\,a(\ell_m^{(z)})$ be the diagonal matrix of attention gains, with one gain for each inducing point.

The Gaussian likelihood treats the rehearsal targets at different inducing points as conditionally independent given their function values. Its expected log-likelihood therefore depends on $\mathbf{S}$ only through the marginal variances $S_{mm}$. Weighting Eq.~\ref{eq:gauss_ell} by the attention gains gives $\sum_m a(\ell^z_m)S_{mm}=\mathrm{tr}(\mathbf{A}\mathbf{S})$ and
\begin{equation}
\begin{aligned}
-\frac{1}{M}\sum_{m=1}^{M} a(\ell^z_m)\,
\E_q\bigl[\log p(t_m\mid f(\bz_m))\bigr]
&=\frac{1}{2\sigma^2 M}(\bm-\mathbf{t})^\top\mathbf{A}(\bm-\mathbf{t})\\
&\quad+\frac{1}{2\sigma^2 M}\mathrm{tr}(\mathbf{A}\mathbf{S})
+\text{const}.
\end{aligned}
\label{eq:maint_nll}
\end{equation}

The KL term of Eq.~\ref{eq:elbo_maint} is $\beta$ times Eq.~\ref{eq:kl_gauss},
\begin{equation}
\begin{aligned}
\beta\,\KL\bigl(q(\bu)\,\|\,p(\bu)\bigr)
&=\frac{\beta}{2}(\bm-\mu_0\mathbf{1})^\top\mathbf{K}_{ZZ}^{-1}(\bm-\mu_0\mathbf{1})
+\frac{\beta}{2}\mathrm{tr}(\mathbf{K}_{ZZ}^{-1}\mathbf{S})\\
&\quad-\frac{\beta}{2}\log|\mathbf{S}|
+\text{const}.
\end{aligned}
\label{eq:maint_kl}
\end{equation}

With the rehearsal target $\mathbf{t}$ held fixed, the sum of~\eqref{eq:maint_nll} and~\eqref{eq:maint_kl} separates into terms involving the mean $\bm$ and terms involving the covariance $\mathbf{S}$. We can therefore differentiate the objective with respect to each quantity separately. The mean-dependent terms are
\begin{equation}
\mathcal{L}_{\mathrm{maint}}(\bm)
=\frac{1}{2\sigma^2 M}(\bm-\mathbf{t})^\top\mathbf{A}(\bm-\mathbf{t})
+\frac{\beta}{2}(\bm-\mu_0\mathbf{1})^\top\mathbf{K}_{ZZ}^{-1}(\bm-\mu_0\mathbf{1})
+\text{const}.
\label{eq:maint_quad}
\end{equation}

Differentiating the mean-dependent objective with respect to $\bm$ gives
\begin{equation}
\nabla_{\bm}\mathcal{L}_{\mathrm{maint}}
=\frac{1}{\sigma^2 M}\mathbf{A}(\bm-\mathbf{t})
+\beta\,\mathbf{K}_{ZZ}^{-1}(\bm-\mu_0\mathbf{1})
=\mathbf{W}\bm-\mathbf{b},
\label{eq:maint_grad}
\end{equation}
Updating the mean in the direction of the negative gradient gives the recurrent equation:
\begin{equation}
\tau\,\dot{\bm}=-\mathbf{W}\bm+\mathbf{b},\qquad
\mathbf{W}=\frac{1}{\sigma^2 M}\mathbf{A}+\beta\,\mathbf{K}_{ZZ}^{-1},\qquad
\mathbf{b}=\frac{1}{\sigma^2 M}\mathbf{A}\,\mathbf{t}+\beta\,\mathbf{K}_{ZZ}^{-1}\mu_0\mathbf{1}.
\label{eq:circuit}
\end{equation}
Here $\mathbf{A}=\mathrm{diag}\,a(\ell^z_m)$ is the diagonal gain matrix from Eq.~\ref{eq:attention}; $\mathbf{W}$ determines recurrent interactions, and $\mathbf{b}$ is the drive from rehearsal targets and the prior mean. The rehearsal target is refreshed after each step, $\mathbf{t}\leftarrow\E_q[f(\bZ)]$, and the derivative~\eqref{eq:maint_grad} treats $\mathbf{t}$ as constant.

The covariance-dependent terms do not depend on the rehearsal target $\mathbf{t}$. Adding the variance term of~\eqref{eq:maint_nll} to the terms in~\eqref{eq:maint_kl} that depend on $\mathbf{S}$ gives
\begin{equation}
\begin{aligned}
\mathcal{L}_{\mathrm{maint}}(\mathbf{S})
&=\frac{1}{2\sigma^2 M}\mathrm{tr}(\mathbf{A}\mathbf{S})
+\frac{\beta}{2}\mathrm{tr}(\mathbf{K}_{ZZ}^{-1}\mathbf{S})
-\frac{\beta}{2}\log|\mathbf{S}|\\
&=\frac12\mathrm{tr}\!\left(
\Bigl(\frac{1}{\sigma^2 M}\mathbf{A}+\beta\,\mathbf{K}_{ZZ}^{-1}\Bigr)\mathbf{S}
\right)
-\frac{\beta}{2}\log|\mathbf{S}|\\
&=\frac12\mathrm{tr}(\mathbf{W}\mathbf{S})-\frac{\beta}{2}\log|\mathbf{S}|,
\end{aligned}
\label{eq:maint_S}
\end{equation}
with the same $\mathbf{W}$ as in~\eqref{eq:circuit}. Encoding and maintenance have related quadratic forms, but their likelihood projections, normalization factors, and KL coefficients differ.
For symmetric $\mathbf{W}$ and $\mathbf{S}$, the derivative is
\begin{equation}
\frac{\partial\mathcal{L}_{\mathrm{maint}}}{\partial\mathbf{S}}
=\frac12\mathbf{W}-\frac{\beta}{2}\mathbf{S}^{-1}.
\label{eq:maint_S_grad}
\end{equation}
Setting the covariance gradient to zero yields the stationary covariance $\mathbf{S}^*=\beta\,\mathbf{W}^{-1}$. Gradient descent on~\eqref{eq:maint_S} follows this derivative. The common factor of one half in the gradient can be absorbed into the time constant $\tau$. The resulting continuous-time gradient update is
\begin{equation}
\tau\,\dot{\mathbf{S}}=-(\mathbf{W}-\beta\,\mathbf{S}^{-1}),
\qquad
\mathbf{S}^*=\beta\,\mathbf{W}^{-1}.
\label{eq:circuit_S}
\end{equation}

Increasing the diagonal entries of $\mathbf{A}$ near the cue increases the corresponding diagonal entries of $\mathbf{W}$ and reduces the corresponding stationary marginal variances in $\mathbf{S}^*$ when the other quantities are fixed. The simulated finite delay need not reach this stationary covariance.

The mean and covariance can respond differently to a cue. When the rehearsal target equals the current mean, $\mathbf{t}=\bm$, the attention-weighted rehearsal term contributes nothing to the mean gradient (Eq.~\ref{eq:maint_grad}), which then contains only the prior-dependent term. Attention still enters the covariance update through $\mathbf{W}$ (Eq.~\ref{eq:circuit_S}). A cue can therefore reduce the variance of the cued strengths without changing their mean, consistent with the simulated dissociation between posterior mean and variance (Fig.~\ref{fig:retrocue}k,l).

\subsubsection*{S3.3 Retrieval}

The maintained mean and covariance $(\bm,\mathbf{S})$ define a posterior over memory strengths. The circuit uses the same retrieval rule as the behavioral simulations: draw strengths jointly across colors and occupied locations, subtract the global sampled minimum to obtain nonnegative responses, and apply spatial divisive normalization as specified in S1.5. The report is $\hat c=\arg\max_c r(c)$. Sampling and normalization are specified readout operations; they are not derived as additional recurrent dynamics from the variational objective.

\subsubsection*{S3.4 Dependence of the recurrent weights on the kernel}

\begin{figure}[htbp]
    \centering
    \includegraphics[width=\textwidth,height=0.8\textheight,keepaspectratio]{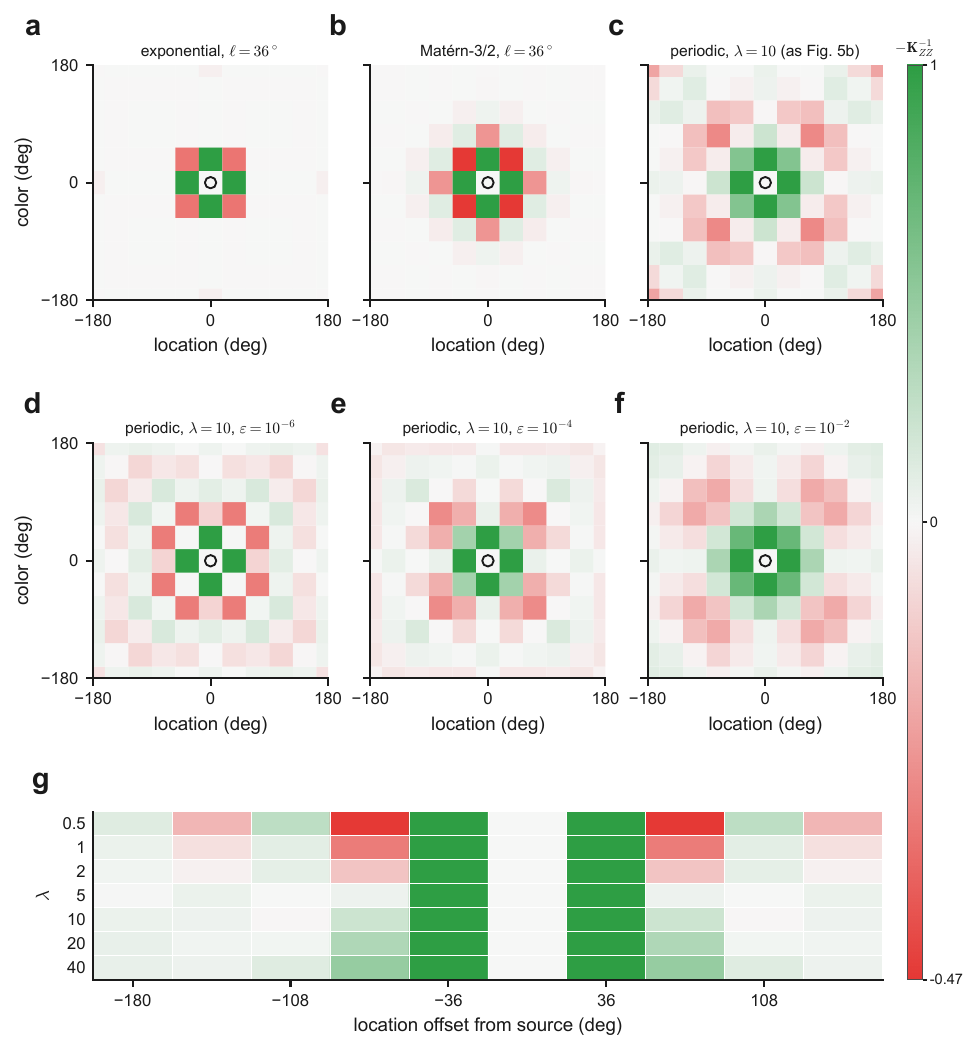}
    \caption{\textbf{Kernel smoothness and conditioning shape the derived connectivity.}
    \textbf{(a--c)} Outgoing weights $-(\mathbf{K}_{ZZ}+\varepsilon\mathbf{I})^{-1}$ from the unit at the origin (open circle; self-coupling removed) on the $10\times10$ inducing grid with $\varepsilon=10^{-3}$, for an exponential kernel (a) and a Mat\'ern-3/2 kernel (b), both with lengthscale $36^\circ$ (one grid step), and for the periodic kernel with $\lambda=10$ used in the paper (c; as in Fig.~\ref{fig:circuit}b). Each is a product kernel over location and color. Weights are scaled by their largest absolute entry; green and red indicate excitation and inhibition.
    \textbf{(d--f)} The periodic kernel with $\lambda=10$ at $\varepsilon=10^{-6}$ (d), $10^{-4}$ (e) and $10^{-2}$ (f).
    \textbf{(g)} Weights along the location axis at the source color for the periodic kernel with $\lambda$ from 0.5 to 40 ($\varepsilon=10^{-3}$).}
    \label{fig:kernel_precision}
\end{figure}

The prior-dependent recurrent weights are the off-diagonal entries of $-\mathbf{K}_{ZZ}^{-1}$, so their pattern depends on the kernel (Supplementary Fig.~\ref{fig:kernel_precision}). Each panel of the figure shows the weights from one unit to all other units on the $10\times10$ inducing grid of Fig.~\ref{fig:circuit}b. Numerical inversion of $\mathbf{K}_{ZZ}$ requires a small jitter $\varepsilon$ on its diagonal, so the panels show $-(\mathbf{K}_{ZZ}+\varepsilon\mathbf{I})^{-1}$, with the self-coupling removed and weights scaled by their largest absolute entry.

The weight pattern also depends on $\varepsilon$ when $\mathbf{K}_{ZZ}$ is nearly singular. The periodic kernel used in the paper ($\lambda=10$) is nearly singular on this grid: the smallest eigenvalue of its one-dimensional kernel matrix is $5\times10^{-8}$. Components of $\mathbf{K}_{ZZ}$ with eigenvalues below $\varepsilon$ are set by the jitter, so the weights change with $\varepsilon$ (Supplementary Fig.~\ref{fig:kernel_precision}d--f). Fig.~\ref{fig:circuit}b uses $\varepsilon=10^{-3}$.

\end{document}